\documentclass[12pt]{article}
\input{amssymb.sty} 

\catcode`\@=11
\def\numberbysection{\@addtoreset{equation}{section}
        \def\theequation{\thesection.\arabic{equation}}}
\numberbysection

\newcommand{\be}{\begin{equation}}
\newcommand{\ee}{\end{equation}}
\newcommand{\bea}{\begin{eqnarray}}
\newcommand{\eea}{\end{eqnarray}}

\def\rest#1{{\langle #1 \rangle}}
\begin{document}
\pagenumbering{alph}
\setcounter{page}{0}
\def\eps{{\epsilon}}
\def\Z{{\mathbb Z}}
\def\N{{\mathbb N}}
\def\R{{\mathbb R}}
\def\Q{{\Bbb Q}}
\def\C{{\Bbb C}}
\def\i{{\rm i}}
\def\cqfd{\hskip 2truemm \vrule height2mm depth0mm width2mm}
\def\boxit#1{\vbox{\hrule\hbox{\vrule{#1}\vrule}\hrule}}
\def\splus{\,\,{\boxit{$+$}}\,\,}
\def\la{\lambda}
\def\si{\sigma}
\def\Dm{{\frak D}_m}
\def\c{{\Bbb Z}}
\def\T{{R}}

\vskip 2truecm

\centerline{{\huge Finite Group Modular Data}}

\vskip 2truecm
\centerline{{\large Antoine Coste}}\smallskip
\centerline{{\it CNRS\footnote{ UMR 8627 } Laboratory of Theoretical Physics}}
\centerline{{\it building 210,  University of Orsay}}
\centerline{{\it 91405 Orsay cedex, France }}
\centerline{{\tt Antoine.Coste@th.u-psud.fr }}
\vskip 1truecm
\centerline{{\large Terry Gannon}}\smallskip
\centerline{{\it Department of Mathematical Sciences,}}
\centerline{{\it University of Alberta}}
\centerline{{\it Edmonton, Canada, T6G 2G1}}
\centerline{{\tt tgannon@math.ualberta.ca}}
\vskip 1truecm
\centerline{{\large Philippe Ruelle\footnote{Chercheur qualifi\'e FNRS}}}
\smallskip
\centerline{{\it Institut de Physique Th\'eorique}}
\centerline{{\it Universit\'e catholique de Louvain}}
\centerline{\it B-1348 \hskip 0.5truecm Louvain-La-Neuve, Belgium}
\centerline{\tt ruelle@fyma.ucl.ac.be}

\vskip 1.5truecm

\centerline{{\bf Abstract}}\vskip .3truecm

In a remarkable variety of contexts appears the modular data associated to
finite
groups. And yet, compared to the well-understood affine
algebra modular data, the general properties of this finite group
modular data has been poorly explored. In this paper we undergo such a
study. We identify some senses in which the finite group data is
similar to, and different from, the affine data. We also consider the
data arising from a cohomological twist, and write
down, explicitly in terms of quantities associated directly with the
finite group, the modular $S$ and $T$ matrices for a {\it general} twist,
for what appears to be the first time in print.
\vfill\eject

\section{Mathematical and Physical Origin}
\pagenumbering{arabic}

Throughout this paper, let $G$ denote any finite group (good references
to finite group and character theory are provided by \cite{K1,K2,BZ}).

Physically, the modular data we will describe in the next section arise in
several ways. It is a (2+1)-dimensional Chern-Simons theory with finite
gauge group $G$ \cite{DW,FQ}. As Witten showed, 3-dimensional topological
field theory corresponds to 2-dimensional conformal field theory
(CFT), and  the corresponding CFTs
here are orbifolds by symmetry group $G$ of a holomorphic CFT
\cite{DVVV,DW,DPR,KT} (e.g. the $E_{8}$ level 1 WZW orbifolded by any
finite subgroup
of the compact simply-connected Lie group $E_8(\R)$).
This CFT incarnation is important to us, as it provides some motivation
for specific investigations we will perform. Nevertheless, both of
these incarnations are probably
of direct value only as toy models.

Note that more generally, to each $G$ we will  obtain a finite-dimensional
representation of the mapping class group  $\Gamma_{g,n}$ for the genus $g$
surfaces with $n$ punctures \cite{B2}. In this paper though we will consider
only the modular group SL$_2(\Z)$, corresponding to $\Gamma_{1,0}$,
and in particular the matrices $S$ and $T$.

It is clear that many different CFTs can realise the same modular data
-- e.g.\ all 71 of the $c=24$ holomorphic theories have
$S=T=(1)$. So any given $G$ will correspond to several different
CFTs. Likewise, all we can say about the central charge $c$ associated
to $G$ is that it will be a positive multiple of 8.

Associated with the RCFT we expect to have some sort of quantum-group which
captures the modular data and representation theory of the chiral algebra
(so e.g.\ the fusion coefficients  are tensor product coefficients of
irreducible modules). This was done for arbitrary finite $G$ in \cite{DPR},
and is the quantum-double of the group algebra $\C [G]$ (see also
\cite{M,B3,KSSB}).

There should also be a vertex operator algebra (VOA) interpretation to this
data, assigning
a VOA to each
finite group. One way to do this is to start with the VOA
associated with any even self-dual Euclidean lattice $\Lambda$ for
which $G\le {\rm Aut}\, \Lambda$. Orbifolding it by $G$ should yield our
data (with the value $c={\rm dim}\,\Lambda$). For example, any finite
subgroup of SU$_2(\C)$ or SU$_3(\C)$ works with the $\Lambda=E_8$ root
lattice. By Cayley's Theorem, such a lattice $\Lambda$ can always be found
for a given group $G$: e.g.\ embed $G$ in some ${\frak S}_n$ and take
$\Lambda$ to be the orthogonal direct sum of $n$ copies of $E_8$. In
\cite{DM,M}, this VOA interpretation is addressed.

Although these holomorphic orbifolds are perhaps too artificial to be of
direct interest, it can be expected that they provide a good hint of the
behaviour of more general orbifolds. Indeed this is
the case -- e.g.\ they can be seen in the theory of permutation orbifolds
\cite{B1}.

Perhaps the most physical incarnation of this modular data is in
(2+1)-dimensional quantum field theories where a continuous gauge group has
been spontaneously broken to a finite group \cite{BDW}. Non-abelian anyons
(i.e.\ particles whose statistics are governed by the braid group rather
than the symmetric group) arise as topological excitations. The effective
field theory describing the long-distance physics is governed by the
quantum-group of \cite{DPR}. Adding a Chern-Simons term corresponds
to the cohomological twist to be discussed shortly.

Actually, this modular data arose originally in mathematics, in an important
but technical way in Lusztig's determination of the irreducible characters
of the finite groups of Lie type \cite{L1,L2}. In describing some of
these, the so-called `unipotent' characters, he was led to consider this
modular
data for the groups $G\in\{\Z_2 \  \times\cdots\times\Z_2 \  ,
{\mathfrak S}_3\ ,{\mathfrak S}_4\   ,{\mathfrak S}_5\  \}$.
For example, ${\mathfrak S}_5$ arises in groups of type $E_8$. Our primary
fields $\Phi$ parametrise the unipotent characters associated to a given
2-sided cell in the Weyl group. \cite{L1} interprets the fusion algebra as
the Grothendieck ring for $G$-equivariant vector bundles.
This Lusztig interpretation is significant as it indicates the richness of
the purely group theoretic side of the data we explore below.

An intriguing application to quantum computing\footnote{ We thank
  Sander Bais for informing us of this. } has been suggested by
Kitaev \cite{ki}. One of the main challenges of actually implementing an effective quantum
computer is decoherence: interaction  with
  the environment makes quantum superpositions very unstable. The
  standard approaches to this involve quantum error correction, but
  Kitaev's proposal is to incorporate into the hardware the
  non-abelian anyons of \cite{BDW}. The resulting computer would operate
  quite robustly amidst localised disturbances.

{}From many of these approaches, we also expect to have
characters ch$_a(\tau)$ which realise this modular data as in (2.1) and (2.2)
below. These appear to be given in \cite{KT}, at least when $G$ is a
subgroup of $E_8(\C)$ (see Theorem 4.3 there).

In summary, what we get is a set of modular data (i.e. matrices $S$ and $T$)
for any choice of finite group $G$. Now, much information about a group can
be recovered easily from its character table: e.g. whether it is  abelian,
simple, solvable, nilpotent, ... (see e.g. Ch.22 of \cite{K1}).
For instance, $G$ is simple iff for all irreducible $\chi\not\equiv 1$,
$\chi(a)=\chi(e)$ only for $a=e$.
Thus it can be expected that our modular data, which probably includes the
character table, should tell us a lot about the group, i.e. be sensitive to
a lot of the group-theoretic properties of $G$.

It does not appear to be known yet whether the modular data associated
to distinct $G$ will always be distinct. 
 A tempting guess would be that this modular data is a
function of the group algebra, but that it
succeeds in distinguishing groups which have different character tables.
 An easy result along these lines is: Given
any groups $G$ and $H$ with equal matrix $S$, if $G$ is abelian then $G\cong
H$. (The character table of an abelian group also uniquely determines
it.) The non-abelian order-8 groups ${\frak D}_4$ and ${\frak Q}_4$
have identical character tables, but their matrices $S$ and $T$ are both
different. More generally, we will see below that the matrices $S$
for the groups with at most 50 primary fields are all distinct.

There are two ways to generalise this data. One is by a `twisting' by some
element of a cohomology group \cite{DW} (see also \cite{DVVV}).
We will look at
this twisting in Section 5. This twisting is rather interesting, because
completely analogously one can speak of the twisting of modular data
associated to Lie groups: the cohomology
group there is $\Z$, and `twisting' by cohomology class $[k]$ gives
WZW at level $k$ for that Lie group ! So the untwisted finite group modular
data  is `level 0' in this sense. On the other hand, the WZW models
at level 0 are trivial. Another difference with WZW is that the cohomology
group here is  finite, so there are only finitely many
different possible `levels' to this finite group data.

The twisting was incorporated in the quantum-group picture in \cite{DPR}.
How to obtain topological (e.g.\ oriented knot) invariants from this
twisted data
was explained in \cite{AC}. It turns out that these knot invariants are
functions of the ``knot group'' (i.e.\ the fundamental group of the
complement of the knot). Though non-isotopic knots can have the same
knot group, this does not imply  that
these  finite group topological invariants are
uninteresting. For instance \cite{Ku}, these invariants can distinguish
a knot from its inverse (i.e. the knot with opposite orientation),
unlike the more familiar topological invariants coming from affine
algebras (or quantum groups $U_q({\frak g})$).

This twisting is reminiscent, though independent,  of
{\it discrete torsion} \cite{V}.
The twist changes the modular data, while discrete torsion changes the
modular invariant partition function but leaves unchanged the 
modular data. We use discrete torsion in \S4.1.

A second way to generalise this data is suggested by the subfactor
interpretation of RCFT. There is a von Neumann subfactor and from this
a fusion ring,
associated to any group-subgroup pair $G\le H$ \cite{KMY}. Our data
comes from the diagonal embedding of $G$ in $G\times G$. The fusion ring
 for generic pairs $G\le H$ is non-commutative (so e.g. Verlinde's
formula will not work), but it is commutative for much more than just
$G\le G \times G$ and so we can expect here a significant generalisation
of this finite
group data corresponding in some way to physics. To our knowledge,
such a physical interpretation has not been developed.

The finite group modular data behaves very differently from the affine
modular data. Yet finite group theory is certainly richer than
nontwisted affine Lie theory, so its modular data should be explored. It
 should  be interesting both from the group
theory and RCFT sides. In particular, it should be interesting to RCFT for
the same reasons as the affine data: Affine modular data is regarded
as physically interesting primarily because it serves as a toy model, and
because of the GKO coset construction; similarly, the finite
group modular data is equally a toy model, and is involved in the orbifold
construction.

As always, there are unexpected surprises. One of the differences we
have found between the finite group data and the affine data is the
large number of modular invariants for the former -- in contrast the
affine modular invariants are surprisingly rare and generally are associated
with Dynkin diagram symmetries. There is one family of exceptions to
this rule: the orthogonal algebras at level 2 have a rich collection
of modular invariants \cite{G}. This was rather mysterious. But what we
find is that the only overlap between the affine data and
nonabelian finite group data occurs with the orthogonal algebras at
level 2 ! This clearly provides a partial resolution of that mystery.

Our main intention with this paper is to make the mathematical
physics community more familiar with this modular data, by developing
and illustrating some
of the basic ideas and constructions. As a large and relatively unexplored
class of RCFTs, studying it can correct some of the prejudices we have
developed  with our rather artificial preoccupation with the affine data.
Finally,
this material  suggests the thought: ``There's some kind of
analogy relating affine algebras and finite
groups'' -- after all, both are directly associated to topological groups
-- and we try to work out some of its aspects. See the Table below. Of
course there are differences as well as similarities, and identifying these
will help us develop a better understanding of generic modular data.

\bigskip\centerline{{\bf List of Notation}}
\bigskip

\begin{tabular}[t]{ll}
$G$ & a group with finitely many elements \\
${\Bbb N}$ & the nonnegative integers \\
ch$_a$ & RCFT characters, see e.g.\ (2.1), (2.8) \\
$\xi_m$ & the $m$-th root of unity $\exp[2\pi\i/m]$ \\
${\Bbb Z}_n$ & the cyclic group (the integers modulo  $n$)\\
${\frak S}_n$ & the symmetric group of order $n!$ \\
${\frak A}_n$ & the alternating group of order $n!/2$ \\
\end{tabular}

\begin{tabular}[t]{ll}
${\frak D}_n$ & the (binary) dihedral group (3.3) of order $2n$ \\
${\frak Q}_{2n}$ & the (generalised) quaternion group (3.4) of order $4n$ \\
$e$ & the identity element in the group \\
$|G|$ & the order (cardinality) of $G$ \\
$K_a$ & the conjugacy class of $a$, i.e.\ $\{gag^{-1}\,|\,g\in G\}$ \\
$\T$ & a set of representatives for each conjugacy class $K_a$\\
$\chi,\psi,\ldots$ & irreducible characters\\
Irr$(G)$ & the set of all irreducible characters of $G$ \\
$G(a,b)$ & see below (2.12) \\
deg$(\chi)$ & dimension of  corresponding representation \\
$C_G(a)$ & the centraliser of $a$ in $G$, i.e.\ $\{g\in G\,|\,ga=ag\}$ \\
$\Phi$ & set of ``primary fields'' $(a,\chi)$ where
$\chi\in{\rm Irr}(C_G(a))$ \\
$e(G)$ & the exponent of $G$ (smallest $n$ such that $g^n=e$ for all $g\in
G$) \\
$Z(G)$ & the centre of $G$ \\
$G'$ & the {commutator subgroup} $\langle ghg^{-1}h^{-1}\rangle$ \\
$k(G)$ & the class number of $G$, i.e.\ the number of conjugacy classes \\
$H \leq G$ & $H$ is a subgroup of $G$ \\
$H\triangleleft G$ & $H$ is a normal subgroup of $G$ \\
$\chi_H^G$ & the induced character of $\chi\in{\rm Irr}(H)$,
where $H\le G$ \\
$M(G)$ & the Schur multiplier $H^2(G,U(1))$ \\
$\beta$ & a $U(1)$-valued 2-cocycle \\
$r(G,\beta)$ & the number of inequivalent projective
$\beta$-representations of $G$\\
$\beta$-Irr$(G)$ & the set of all irreducible projective $\beta$-characters
of $G$ \\
$\tilde\chi,\tilde\psi,\dots$ & irreducible projective characters \\
$\alpha$ & a 3-cocycle in $H^3(G,U(1))$ \\
CT & a 3-cocycle $\alpha$ such that all 2-cocycles $\beta_a$ are
coboundaries (see \S5.2)\\
${ S}^\alpha$, ${ T}^\alpha$ & the modular data twisted
by a 3-cocycle $\alpha$\\
${\Bbb F}_q$ & the finite field with $q$ elements \\
\end{tabular}

\vskip 1truecm
\bigskip
\centerline{{\bf The affine algebra versus finite group analogy}}
\bigskip

\begin{tabular}[t]{lll}
primary fields in $\Phi$ & $r$-tuples $\la\in P_+(X_{r,k})$ &  pairs $(g,
\chi)$ \\
level & a number $k\in\N$ & a twist $\alpha\in H^3(G,U(1))$ \\
simple currents & extended Dynkin diagram & $Z(G)\times G/G'$ (untwisted
case)\\
 & \ \ \ \ symmetry (except $\hat E_{8,2}$) & \\
conjugation & unextended Dynkin symmetry & involutive Galois permutations\\
conformal embedding & $X_{r,k}\subset Y_{s,1}$ &
perhaps $H\triangleleft G$ \\
central charge & $c={k\,{\rm dim}\ X_r\over k+h^\vee}$ & $c\in 8\Z $ \\
Chern-Simons theory&gauge group is Lie gp &gauge group is finite group \\
CFT&WZW model&holomorphic orbifold \\
quantum-group&$U_q(X_r)$, $q^N=1$&quantum double ${D}^\alpha(G)$ \\
VOA & Frenkel-Zhu & Dong-Mason \\
rank-level duality & $\widehat{\rm su}(n)_k\leftrightarrow\widehat{\rm
su}(k)_n, \ldots$  &
electric/magnetic duality(??)    \\
\end{tabular}


\bigskip
\section{Untwisted modular data}

\subsection{Modular data for any RCFT}

In this paper we explore what we call the modular data of $G$, from the
perspective of RCFT. This subsection is intended as a quick review of
the basic RCFT data (see e.g. \cite{DMS} for a more comprehensive
treatment), and we end it with a brief description of the modular data
associated with affine algebras.

Let $\Phi$ denote the (finite) set of primary fields in the RCFT. One
of these primaries, the ``vacuum'' `0', is privileged. By the {\it modular
data} we mean the modular matrices $S$ and $T$, whose entries $S_{ab}$ and
$T_{ab}$ are parametrised by $a,b\in \Phi$. These matrices are unitary, $T$
is diagonal,
$S$ is symmetric, and together they define a representation of SL$_2(\Z)$
realised by the modular action on the RCFT characters ch$_a(\tau)$:
\bea
{\rm ch}_a(-1/\tau)=&\sum_{b\in\Phi}S_{ab}\,{\rm ch}_b(\tau),\\
{\rm ch}_a(\tau+1)=&\sum_{b\in\Phi}T_{ab}\,{\rm ch}_b(\tau),
\eea
so $(ST)^3=S^2=:C$ commutes with $S$ and $T$ and is a permutation
matrix called {\it charge conjugation}. Note that $S^*=SC$ and $C0=0$.
These ch$_a$ are assumed to be linearly independent and in general
will depend on other variables than $\tau$, but it is conventional to
write only $\tau$.

All RCFTs in this paper are {\it unitary}. This implies that the entries
$S_{a0}=S_{0a}$ will all be strictly positive. The ratio ${S_{a0}\over
S_{00}}$ is called the {\it quantum dimension} of $a$.
The {\it fusion coefficients}
$N_{ab}^c \in \N$ of the theory can be obtained from $S$ using {\it
Verlinde's formula}
\be
N_{ab}^c=\sum_{d\in\Phi}{S_{ad}\,S_{bd}\,S_{cd}^*\over S_{0d}}.
\ee

A {\it simple current} \cite{SY} can be defined as any $j\in\Phi$ with
quantum-dimension 1: i.e.\ $S_{j0}=S_{00}$. To any simple current $j$ is
associated a ``charge'' $Q_j:\Phi\rightarrow\Q$, an integer $R_j$,
 and a permutation $J$ of $\Phi$, such that $J0=j$,
\bea S_{Ja,b}&=&\exp[2\pi\i \,Q_j(b)]\,S_{ab},\\
T_{Ja,Ja}T^*_{aa}&=&\exp[2\pi\i\,(R_j\,{\textstyle {n-1\over
2n}}-Q_j(a))],\\ N_{j,b}^c&=&\delta_{c,Jb},\eea
where $n$ in (2.5) is the order of $J$. Note that $nQ_j\in\Z$.
For instance $j=0$ is a simple current, corresponding to the identity
permutation. Simple currents define an abelian group given by composition
of the corresponding permutations.

The matrix $S$ also obeys another important symmetry, called {\it
Galois} \cite{CG}. In particular, the entries $S_{ab}$ will lie in some
cyclotomic field $\Q(\xi_m)$, where $\xi_m$ throughout this paper denotes
the $m$th
root of unity $\exp[2\pi\i/m]$. This means that each $S_{ab}$ can be
written as a
polynomial in $\xi_m$ with rational coefficients. The Galois group
Gal$(\Q(\xi_m)/\Q)$ consists of all automorphisms of the field
$\Q(\xi_m)$, and is isomorphic to the multiplicative group $\Z_m^\times$
of integers coprime to $m$. In particular, to any $\ell\in\Z_m^\times$
we get an automorphism $\si_\ell\in{\rm Gal}(\Q(\xi_m)/\Q)$ sending
$\xi_m$ to $\xi_m^\ell$. Applying it to entries of $S$, we get
\be
\si(S_{ab})=\eps_\si(a)\,S_{\si(a),b}=\eps_\si(b)\,S_{a,\si(b)},
\ee
where each $\eps_\si(a)$ is a sign $\pm 1$, and where $a\mapsto
\si(a)$ is a permutation of $\Phi$. The simplest example of this
Galois action is charge conjugation $C$: $\si_{-1}$ is complex
conjugation; the corresponding $\eps_{\si_{-1}}$ is identically +1,
while the permutation $a\mapsto \si_{-1}a$ is given by $C$.

The Galois and simple current permutations respect each other:
$\si_\ell(Ja)=J^\ell(\si_\ell(a))$. Also, $\eps_\si(Ja)=\eps_\si(a)$
and $Q_j(\si_\ell(a))\equiv\ell \,Q_j(a)$ (mod 1).

The one-loop partition function ${\cal Z}(\tau)$ of the RCFT is a
modular invariant sesquilinear combination of RCFT characters:
\be
{\cal  Z}(\tau)=\sum_{a,b\in\Phi}M_{ab}\,{\rm ch}_a(\tau)\,{\rm ch}_b(\tau)^*.
\ee
Modular invariance means that the coefficient matrix $M$ commutes with
$S$ and $T$. In addition, $M_{00}=1$, and each coefficient is a
non-negative  integer: $M_{ab} \in \N$. Any such ${\cal Z}$ or $M$
is called a {\it physical invariant} or {\it modular invariant} (such an
$M$ may or may not be
realised as a partition function for an RCFT). $M=I$ is always a
physical invariant. For a given choice of modular data, there will only
be finitely many physical invariants.

A special family of physical invariants are the {\it automorphism
invariants}, where $M$ is a permutation matrix. An example is $M=C$. 
The product $MM'$ of any automorphism invariant $M$ with
any physical invariant $M'$ will also be a physical invariant.

One of the uses of both simple currents and this Galois action is that
they can be used to construct physical invariants. The simplest
construction (originally due to \cite{Be} but since generalised
considerably, see
\cite{KS} and references therein) starts with any simple current $j$, with
order
$n$ say. Then
\be
M(j)_{ab}=\sum_{i=1}^n\delta_{J^ia,b}\,\delta^1(Q_j(a) + {i\over  2n}R_j),
\ee
where $\delta^1(x)=1$ if $x\in\Z$ and $=0$ otherwise. For instance the
choice $j=0$ yields the identity matrix $M(0)=I$. $M(j)$ will be a
physical invariant iff  $T_{jj}
T^*_{00}$ is an $n$th root of 1. $M(j)$ will be an automorphism invariant
iff $T_{jj}T^*_{00}$ is a {\it primitive} $n$th root of 1.

Given any order-two Galois automorphism $\si$ with the properties that
$\si(0)=0$ and $T_{\si a,\si a}=T_{a,a}$, then $M(\si)_{ab}:=\delta_{b,\si a}$
defines an automorphism invariant (this construction was originally
due to \cite{GFSS} but was since generalised). Again, $C$ is an example.

The most familiar example of modular data comes from the
affine nontwisted algebras. The literature on affine modular data is
very extensive (indeed this affine $\leftrightarrow$ finite group
imbalance is a primary motivation for this paper) and it certainly is not
our intention to review it here. See \cite{K,DMS} for a deeper treatment.

Choose any finite-dimensional simple Lie algebra $X_r$, and any {\it
level} $k \in \N$. $X_r^{(1)}=\hat{X}_r$ denotes the infinite-dimensional
nontwisted affine Kac-Moody algebra \cite{K}. Its integrable highest-weight
modules at level $k$ are parametrised by ($r+1$)-tuples
$\la=(\la_0,\la_1,\ldots,\la_r)$, where each $\la_i \in \N$, and
$\sum_{i=0}^ra_i^\vee \la_i=k$. The positive constants $a_i^\vee$ are
called colabels -- e.g.\ for $A_r^{(1)}$ or $C_r^{(1)}$ all $a_i^\vee=1$.
The (finite) collection of these highest-weights is denoted $P_+^k$
and is the set of primary fields $\Phi$. The ``vacuum'' 0 is
$(k,0,0,\ldots,0)$. The matrices $S$ and $T$ are given in Theorem 13.8 of
\cite{K}: $S$
is related to the corresponding Lie group characters at elements of finite
order, while $T$ is related to the eigenvalues of the quadratic Casimir.
A useful parameter is the dual Coxeter number
$h^\vee := \sum_{i=0}^ra_i^\vee$, and a useful $(r+1)$-tuple is the Weyl
vector $\rho:=(1,1,1,\ldots,1)$. The affine fusion coefficients are given
combinatorially by what is usually called the Kac-Walton formula
\cite{K,Wa}, though it has other co-discoverers.

The simplest example is the choice $X_r=A_1$ (i.e. sl$_2(\C)$) for which
$h^\vee=2$. For any level $k \in {\Bbb N}$, one has
$P_+^k=\{(k,0),(k-1,1),\ldots,(0,k)\}$. The modular data is then
\bea
S_{\la\mu}=&\sqrt{2\over k+2}\,\sin(\pi{(\la_1+1)(\mu_1+1)
    \over k+2}),\\
T_{\la\mu}     =&\exp[\pi\i{(\la_1+1)^2 \over 2(k+2)}-
{\pi\i\over 4}]\,  \delta_{\la,\mu }.
\eea

The charge conjugation of affine modular data corresponds to an
order-one or -two symmetry of the unextended Dynkin diagram. For
$A_1^{(1)}$ it is trivial, but for the other $A_r^{(1)}$ it is
non-trivial. The simple currents were classified in \cite{F}, and with one
exception ($E_8^{(1)}$ level 2) correspond
to  symmetries of the extended Dynkin diagram. $A_1^{(1)}$ has one
non-trivial such symmetry, corresponding to the interchange of the
nodes labeled `0' and `1'. This yields a permutation $J$ of $P_+^k$
given by $J(\la_0,\la_1)=(\la_1,\la_0)$, and corresponding to primary
field $j=J0=(0,k)$. Note that $Q_j(\la)=(-1)^{\la_1}$ and
$R_j=k$. Thus $M(j)$ is a physical invariant whenever $k$ is even. It
is an automorphism invariant whenever $k\equiv 2$ (mod 4).

The Galois action can be understood geometrically in terms of the
affine Weyl group \cite{CG}. In general $\si(0)\ne 0$ and the signs
$\eps_\si$ are both positive and negative. The $S$  entries lie in
the cyclotomic field $\Q(\xi_{f\,(k+h^\vee)})$, where $f$ is a number
depending only on the choice of algebra -- e.g.\ $f=4$ for $A_1$, and
$f=3,4,1,4,3$ for $X_r=E_6,E_7,E_8,F_4,G_2$, respectively (we will use
this in the proof of Theorem 1 below). Also, $h^\vee=12,18,30,9,4$ for
those algebras.

A surprising feature is that almost all of the affine physical invariants
can be understood in terms of the symmetries of the extended Dynkin
diagram. For example, the classification for $A_1^{(1)}$ was done in
\cite{CIZ}: apart from $M=M(0)=I$ (for all $k$) and $M=M(j)$ (for all even
$k$), there are only 3 other physical invariants (at $k=10,16,28$).
The physical invariant classification for general affine algebras is
still open; see \cite{G} and references therein for a list of the algebras
and levels for which it has been completed.

\subsection{Finite group modular data}

After this brief review of modular data in general RCFTs, and of affine
modular data as specific examples, we turn to the modular data associated
with a finite group $G$.

Fix a set $\T$ of representatives of
each conjugacy class of $G$. So the identity $e$ of $G$ is in $\T$, and
more generally the centre $Z(G)$ of $G$ is a subset of $\T$.

For any $a\in G$, let $K_a$ be the conjugacy class containing $a$. By
$C_G(a)$ we mean the {\it centraliser} of $a$ in $G$, i.e. the set of all
elements in $G$ which commute with $a$. $C_G(a)$ is a subgroup of $G$,
and in fact $|G| = |K_a| \cdot |C_G(a)|$.

The primary fields of the $G$ modular data are labelled by pairs $(a,\chi)$,
where $a\in \T$, and where $\chi$ is an irreducible character of $C_G(a)$.
We will write $\Phi=\Phi(G)$ for the set of all these pairs. In this set,
the vacuum `0' corresponds to $(e,1)$, with 1 the character of the
trivial representation of $G = C_G(e)$.  It will be convenient at
times to identify $(a,\chi)$ with each $(g^{-1}ag,\chi^g)$, where
$\chi^g(h)=\chi(ghg^{-1})$ is an irreducible character of
$C_G(g^{-1}ag)$.

We set
\bea
S_{(a,\chi),(b,\chi')} &=& {1\over |C_G(a)|\,|C_G(b)|}\sum_{g\in
G(a,b)} \chi(gbg^{-1})^* \, \chi'(g^{-1}ag)^* \label{uS1} \\
&=& {1\over |G|}\sum_{g\in K_a,h\in K_b\cap C_G(g)}
\chi(xhx^{-1})^* \, \chi'(ygy^{-1})^*, \label{uS2} \\
T_{(a,\chi),(a',\chi')} &=& \delta_{a,a'}\delta_{\chi,\chi'}{\chi(a)\over
\chi(e)}, \label{uT}
\eea
where $G(a,b)=\{g\in G\,|\,agbg^{-1}=gbg^{-1}a\}$, and where $x,y$ are
any solutions to $g=x^{-1}ax$ and $h=y^{-1}by$. The equivalence of
(\ref{uS1}) and (\ref{uS2}), and the fact that (\ref{uS2}) does not depend
on the choice of $x,y$, are easy arguments.

Note that the strange set $G(a,b)$ is precisely the set of all $g$ for
which $g^{-1}ag \in C_G(b)$ and $gbg^{-1}\in C_G(a)$. If there are no
such $g$, then $G(a,b)$ is empty and the sum (and the matrix entry)
would be equal to 0. A special case of this is Proposition 1(a) below.

$S$ is symmetric and unitary, and gives rise (via Verlinde's formula) to
non-negative integer fusion coefficients. The fusion coefficient
$N_{(a,\chi_1),(b,\chi_2)}^{(c,\chi_3)}$ in fact has algebraic
interpretations. For example, let $\rho_i$ be representations corresponding
to  characters $\chi_i$. Write $T(a)$ for a set of representatives of the left
cosets of $C_G(a)$. Define the space
\be
X=\bigoplus \; (\rho_1x) \otimes (\rho_2y),
\ee
where the sum is over all $x\in T(a)$, $y\in T(b)$ for which $(x^{-1}ax)\,
(y^{-1}by)=c$, and where we interpret `$\rho_1x$', `$\rho_2y$' as belonging to
the induced representations $(\rho_1){C_G(a)}^G$, $(\rho_2)_{C_G(b)}^G$,
respectively (we'll say more on induced representations below).
$X$ carries a representation of $C_G(c)$, using the usual coproduct. The
fusion coefficient equals the multiplicity of $\rho_3$ in $X$ \cite{M}.
(There are
other interpretations of these fusion coefficients -- see e.g.
\cite{L1,DVVV,DPR}.)

Since $a$ is in the centre of $C_G(a)$, $\chi(a)/\chi(e)$ in (\ref{uT})
will be
an $n$th root of 1, where $n$ is the order of
$a$. Because $T_{00}=1$, the given normalisation of $T$ corresponds to the
central charge $c$ being a multiple of 24; we are free to multiply
(\ref{uT}) by any third root of 1, permitting $c$ to be other multiples of
8.

In order to more fully exploit the formula (\ref{uS1}), it is important to
understand the notion of {\it induced character} (or representation) (see
e.g. \cite{BZ}).  Given a representation
$\rho : H \rightarrow W$ of a subgroup $H$ of $G$, of character $\chi$, we
call a representation $\rho':G\rightarrow V$ of $G$ an induced
representation if $V$ equals the direct sum of vector spaces $W_a$, for
each coset $aH\in G/H$, where $W_a$ is defined by $\rho'(aH)W=W_a$. An
induced representation always  exists and is unique up to isomorphism, and
is denoted $\rho_H^G$. In terms of characters, we get the important formula
(somewhat reminiscent of (\ref{uS1})):
\be
\chi_H^G(g)={1 \over |H|}\sum_{{a\in G\atop a^{-1}ga\in H}}
\chi(a^{-1}ga). \label{ind}
\ee

Suppose that $\chi'\in{\rm Irr}(C_G({b}))$ is the
restriction to $C_G(b)$ of some character $\overline{\chi}'$ defined
on the group $G(a|b):=\langle G(a,b),C_G(b)\rangle$ (this happens
fairly often in practice -- see e.g.\ Ch.27 of \cite{K1} for a
relevant discussion). Then provided $G(a,b)\ne \emptyset$, (\ref{uS1})
collapses to
\be
S_{(a,\chi),(b,\chi')} = {1\over |C_G(b)|}
\,\overline{\chi}'(a)^*\,\chi_{C_G(a)}^{G} (b)^*. \label{rest}
\ee

Equation (\ref{rest}) for instance applies whenever $\chi'$ is identically
1. Another  important instance of (\ref{rest}) is when $z$ lies in the
centre $Z(G)$, since then $C_G(z)=G$ and we get
\be
S_{(a,\chi),(z,\chi')}={\chi(z)^*\over |C_G(a)|}\,\chi'(a)^*. \label{cen}
\ee

For instance, the quantum dimension of
$(a,\chi)\in \Phi$ is
\be
{S_{(a,\chi),(e,1)}\over S_{(e,1),(e,1)}}=|K_a|\,{\rm deg}(\chi),
\label{qdim}
\ee
which amazingly enough is always a positive {\it integer} ! In fact from
(\ref{ind}), (\ref{qdim}) has a simple group-theoretic interpretation: it is
the degree of the induced character $\chi_{C_G(a)}^G$. This integrality is
very unusual, and shows that generically this finite group
modular data is qualitatively different from affine data.

Equation (\ref{rest}) says we can expect many 0's in $S$. Any character
$\chi\in{\rm Irr}(H)$ with degree $\chi(e)>1$ will have a zero -- in
fact $\chi(h)$ will be zero for at least $|Z(H)|\,(\chi(e)^2-1)$ group
elements (see
e.g.\ Ch.\ 23 of \cite{K1}). Hence  the $(z,\chi)$ row will have 0's iff $\chi$
is not of  degree 1. Dually, for any $b\in H$ there will be at least
$k(H)-|C_H(b)|$ different $\chi'\in{\rm Irr}(H)$ with $\chi'(b)=0 $,
where $k(H)$ is the class number of $H$,
 so for instance the $(a,1)$ row will have 0's whenever e.g.\ there are elements in
 $C_G(a)$ whose centraliser in $C_G(a)$ is
 small. This seems to also be
different from the affine case, where 0's are quite rare and fairly
generically tend to be  due to simple current fixed points.

Equation (\ref{cen}) also says
\be
{S_{(e,\chi), (a,\chi')}\over S_{(e,1),(a,\chi')}}=\chi(a)^*. \label{star}
\ee
As long as we could identify by looking at $S$ and $T$ the primaries of the
form $(e,\star)$, then the matrix $S$ will contain the character table of
$G$, and we would know that groups with different character tables would
necessarily have different matrices $S$ and $T$.

{}From (\ref{qdim}), we see that the simple currents are precisely the pairs
$(z,\varphi)$, where  $z$ lies in the
centre $Z(G)$ of $G$, and $\varphi$ is a degree-1 character of $G$.
Thus the group of simple currents is isomorphic to the direct product
$Z(G)\times G/G'$, where $G'$ is the commutator subgroup of $G$.
The simple current $j=(z,\varphi)$ corresponds to permutation
 $J_{(z,\varphi)}(a,\chi)=
(za,\varphi\chi)\in\Phi$, and charge  $e^{2\pi \i Q}
= \varphi(a)^*\chi(z)^*/\chi(1)$,
which is always a root of unity as it should be.
 We get that $M(z,\varphi)$ (in the
notation of the previous subsection) is always a physical invariant; it
is an automorphism invariant iff $z$ and $\varphi$ have the same order.

Generic groups have many simple currents. A group for which $G=G'$ is
called {\it perfect}; a group will have no non-trivial simple currents
iff it is perfect and has trivial centre.  For example this happens
whenever $G$ is non-cyclic simple. All perfect groups with small
orders have been classified \cite{HP}, and using this we can list all
groups $G$ with order $|G| < 688128$ which have no non-trivial simple
currents. The orders under 2000 of these groups are (see also Prop.\ 1(g)):
60, 168, 360, 504, 660, 960 (twice), 1092, 1344 (twice), and 1920.

When (and only when) $G$ is abelian, all primaries will be simple currents,
and hence the modular data will be rather trivial and uninteresting.
This can also happen with the affine algebras: namely, the
simply-laced algebras $A_r^{(1)},D_r^{(1)},E_6^{(1)},E_7^{(1)},E_8^{(1)}$
at level 1.

Charge conjugation takes $(a,\chi)$ to $(a^{-1},\chi^*)$,
where the complex conjugate $\chi^* $ is the character of the
contragredient representation. Now $a^{-1}$ may not lie in our set $\T$
of conjugacy class representatives: recall that by $(a^{-1},\chi^*)$
we really mean $(g^{-1}a^{-1}g,\chi^{g\ *})\in\Phi$ where
$g^{-1}a^{-1}g\in \T$. This is not so trivial as it may seem: if $a$ and
$a^{-1}$ are conjugate, $(a,\chi)$ and $(a^{-1},\chi^*)$ may be
identified even though $\chi$ is complex-valued. An example is ${\frak
S}_3$, $a=(123)$, given below. Note though from (\ref{star}) that the field
$\Q(S)$ contains the field generated over $\Q$ by all character values
$\chi(a)$, $\chi \in {\rm Irr}(G)$, $a\in G$, so if some characters of
$G$ are complex, the $C$ won't be trivial.

The Galois symmetry is also  straightforward. The character values of
any group lie in the cyclotomic field $\Q(\xi_e)$ where $e=e(G)$ is the
exponent of the group (the least common multiple of the orders of all
group elements) -- see Thm.8.7 of \cite{BZ}. Hence the entries of $S$ and
$T$ lie in the cyclotomic field $\Q(\xi_e)$, so the relevant Galois
group is the multiplicative group $\Z_e^\times$. The Galois automorphism
$\sigma_\ell$ takes $(a,\chi)$ to $(a^\ell,\sigma_\ell\circ\chi)$:
$\sigma_\ell
\circ\chi$ is the function obtained by applying $\si_\ell$ to each
complex number $\chi(a)$; it is an irreducible character (of degree equal
to that of
$\chi$) iff $\chi$ is \cite{BZ}. Note another curiousity here: the Galois
parities
$\epsilon_\ell$ are all identically equal to 1 ! This is very
different from the generic affine situation. Note also that every Galois
permutation fixes the vacuum $(e,1)$ (since every quantum dimension is
rational), so large numbers of automorphism invariants will
arise generically: whenever $\ell^2 \equiv 1$
(mod $e(G)$), the permutation $\si_\ell$ on $\Phi$ will
define an automorphism invariant. We will return to this in \S4.1.

Let us collect a few of the observations we have made here.

\smallskip \noindent{{\bf Proposition 1}}. (a) Choose any
$(a,\chi),(a',\chi')\in \Phi$. If the order of $a$ does not divide the
exponent of $C_G(a')$, or if the order of $a'$ does not divide the exponent
of $C_G(a)$, then $S_{(a,\chi),(a',\chi')}=0$.
\begin{itemize}
\item[(b)] The order of $T$ {\it equals} (and not merely divides)  the
exponent of $G$.

\item[(c)] If ${ S}(G)={ S}(H)$, then $|G|=|H|$.

\item[(d)] If $a^{-1}\not\in K_a$, then the charge conjugate
$C(a,\chi)\ne(a,\chi)$ for all $\chi\in{\rm Irr}(C_G(a))$. If
$\chi\in{\rm Irr}(G)$ is not real-valued, then
$C(z,\chi)\ne(z,\chi)$ for all $z\in Z(G)$.

\item[(e)] The quantum dimensions $S_{(a,\chi),(e,1)}/S_{(e,1),(e,1)}$
  are always integers.

\item[(f)] The Galois parities $\epsilon_\si(a,\chi)$ are always +1, and the
vacuum $(e,1)$ is fixed by all $\si$.

\item[(g)] The groups $G\ne \{e\}$ with at most 75 primaries, which have no
 non-trivial simple currents, are the simple groups $G={\frak A}_5,{\rm
   PSL}_2({\Bbb F}_7),{\frak A}_6, {\rm SL}_2({\Bbb
   F}_8),{\rm PSL}_2({\Bbb F}_{11})$, and ${\frak A}_7$, with
 $(|G|,|\Phi|)=(60,22),(168,35),(360,44), (504,74), (660,58)$, and
 (2520,74).

\item[(h)] For any group $G\ne\{e\}$, there will be at least 5
  physical invariants. If $G/Z(G)$ is nonabelian, then there will be
  at least $a+(b+c)^2$ physical invariants, where $a=|Z(G)|\,|G/G'|$,
  $b$ equals the number of subgroups of $Z(G)$, and $c$ equals the
  number of subgroups of $G/G'$.
\end{itemize}\smallskip

The proof for (b) is the following.
Choose any group $H$, and any
 $a\in Z(H)$ with prime-power order $p^m$. Then any irreducible
 representation of $H$ will send $a$ to the multiple of the identity
 matrix by some $p^m$th root of 1, i.e.\ $\chi(a)/\chi(e)$ will be a
 $p^m$th root of 1 for all $\chi\in{\rm Irr}(H)$. Now, that root of 1
 must be primitive for some $\chi$, as otherwise $a^{p^{m-1}}$ and $e$
 would have identical character values.  Applying that fact to
 $H=C_G(a)$ gives us the desired order of $T$. 

The proof of (c) is the comparison of $S_{(e,1),(e,1)}$ for $G$ and $H$.

A constructive proof of (h) is given in \S4.2. Incidentally,
this bound can probably be significantly improved.
$G={\c}_2$, with 6 physical invariants, is probably the lowest number. By
comparison, affine algebras have relatively few physical invariants: e.g.
both $B_{\ell}^{(1)}$ and $C_{\ell}^{(1)}$ at generic levels are expected
to have only 2. \cqfd

We have already remarked on several occasions above, that the finite group
modular data are very different from affine modular data. At this stage (we
will have more to say when we come to general twisted finite group modular
data), this statement can be substantiated by noting that the two sets have
a very small intersection.

\smallskip\noindent{{\bf Theorem 1.}}
Let $S$ and $T$ be the Kac-Peterson matrices corresponding to an affine algebra
$X_r^{(1)}$ at some level $k \geq 1$ (where $X_r$ is simple). Let $G$ be a
finite
group with ${ S}(G) = S$ and  ${ T}(G)=\varphi\,T$ for some third root
$\varphi$ of 1. Then either

\begin{itemize}

\item[{\it (i)}] $(X_r,k)=(E_8,1)$ and $G=\{e\}$, or

\item[{\it (ii)}] $(X_r,k)=(D_{8n},1)$, and $G=\Z_2$.

\end{itemize}

\smallskip
\noindent {\it Sketch of proof}\qquad
Write $n=k+h^\vee$. Recall that there is a Galois automorphism for any
$\ell$ coprime to $fn$, where $f$ for the exceptional algebras was
given in  \S2.1. A consequence of the affine Weyl interpretation of the
affine Galois permutation $\si_\ell$ of $P_+^k$ is that the vectors
$\si_\ell(\la+\rho)$ and $\ell\,(\la+\rho)$ have the same norm mod $2n$.

A nice  way to handle the exceptional algebras is to check that $\ell\rho$
and $\rho$ have the same norm (mod $2n$) for any $\ell$ coming from Galois
(since all Galois automorphisms here will fix the vacuum 0). For
instance, for $E_7$ we get $\rho^2=399/2$, so we see that $2n$ must
divide $(\ell^2-1)\, 399/2$. Now, the ``Definition of 24'' says that
$\ell^2-1$ here can be replaced with 24: more precisely, the gcd of
all numbers $\ell^2-1$, for $\ell$ coprime to $fn$, will equal
gcd$(24,(fn)^\infty)$. Hence, $n$ must divide $2\cdot 3^2\cdot 7\cdot 19$.

Now, $n>h^\vee=18$. Also, if $n$ is too big (i.e.\ if there is an
$\ell$ coprime to $fn$ such that $(h^\vee-1)\ell<n$), then
$\si_\ell(0)$ will equal $(m,\ell-1,\ell-1,\ldots,\ell-1)$, where
$m=n-1-\ell\,(h^\vee-1)$, violating the result that 0 is fixed by all
Galois automorphisms. For $E_7$, this is the condition that no $\ell$,
$1<\ell<{n\over 17}$, can be coprime to $2n$.

We can now write down the possibilities for $n$: they are $19, 21, 38, 42,57,
63$. Subtracting 18 gives the possible levels. $n=19$ fails, since it
would have to correspond to an abelian group with order ${1\over
S_{00}}=\sqrt{2}$.
The remaining 5 possibilities all have non-integral quantum dimensions.

The other exceptional algebras are all done similarly. The only
surviving $n$ for $G_2$ are $n=7,8,14$; for $F_4$ are $n=12,13,18,36,39$; for
$E_6 $ are $n=18,24,26,36,52$; and for $E_8$ are $n=40,48,60,62,80,120 $.
In all these cases, a weight with non-integral quantum dimension is
easily found.

The best way to handle the classical algebras is to compute the
quantum dimension of any weight in the Galois orbit of 0, and
show it must be larger than 1, for some $\si_\ell$.
For now consider $k>1$ for $A_r$ and $C_r$, and $k>2$ for $B_r$ and $D_r$.
Up to a sign, the quantum dimension of $\si_\ell (0)$ for the algebras
$A_r,B_r,C_r,D_r$ is, respectively,
\be
\prod_{a=1}^r \;{\sin(\pi\ell a/n)^{r+1-a}\over\sin(\pi  a/n)^{r+1-a}},
\ee
\be
\prod_{a=0}^{r-1}\;{\sin(\pi\ell\,(a+1/2)/n)\over \sin(\pi\,(a+1/2)/n)}\,
\prod_{b=1}^{2r-2}{\sin(\pi\ell b/n)^{[{2r-b\over 2}]}\over\sin(\pi
  b/n)^{[{2r-b\over 2}]}},\ee
\be
\prod_{a=1}^{r-1}\;{\sin(\pi\ell a/n)^{r-a}\,\sin(\pi\ell\,(a-1/2)/n)^{r-a}
\over \sin(\pi a/n)^{r-a}\,\sin(\pi\,(a-1/2)/n)^{r-a}}\,
\prod_{b=r}^{2r-1}{\sin(\pi\ell b/2n)\over \sin(\pi b/2n)},\ee
\be
\prod_{a=1}^{r-1}\;{\sin(\pi\ell a/n)^{[{2r-a+1\over 2}]}\over \sin(\pi
  a/n)^{[{2r-a+1\over 2}]}}\,\prod_{b=r}^{2r-3}{\sin(\pi\ell
    b/n)^{[{2r-b-1\over 2}]} \over \sin(\pi b/n)^{[{2r-b-1\over 2}]}},
\ee
where $[x]$ here denotes the greatest integer not more than $x$.
The absolute value of each of these is quickly seen to be greater than 1 unless $\ell
\equiv \pm 1$ (mod $n$) (for $A_r$ or $D_r$) or $\ell\equiv \pm 1$
(mod $2n$) (for $B_r$ and $C_r$). This exhausts the
possible $\ell$ coprime to $fn$ only if the Euler totient
$\varphi(n)\le 2$ (for $A_r$ and $D_r$) or $\varphi(2n)\le 2$ (for
$B_r$ and $C_r$). By definition $\varphi(m)$ is the number of $h$,
$1\le h\le m$, coprime to $m$; it is less than 3 only for $m=1,2,3,4,6$.
So only $A_{1}$ level 2, $A_{2}$ level 3, $A_{1}$ level 4, and $A_3$
level 2 survive, but their central charges aren't multiples of 8.

The series $A_r$, $D_{odd}$ and $D_{even}$ at $k=1$ possess only
simple currents (so would have to correspond to an abelian group $G$),
and have  the fusion groups $\Z_{r+1},
\Z_4,\Z_2\times\Z_2$ respectively. Abelian $G$ has fusion group
$G\times G$, so that leaves only $D_{even}$ and $G=\Z_2$. $D_n$ level
1 has $c=n$, concluding the argument. $B_r$ level 1 has only 2
primaries so is covered e.g.\ by Theorem 2 below. The modular data of
$C_r$ level 1 is identical with that of $A_1$ level $r$.

Level 2 for $D_r$
can be handled by requiring $|G|={1\over S_{00}}=2\sqrt{2r}$ and the
quantum dimension ${S_{\Lambda_r,0}\over S_{00}}=\sqrt{r}$ to both be
integers (see \cite{G} for the necessary $S$ entries).
For $B_r$ at level 2, first read off from \cite{G} that $|G|={1\over S_{00}}=
2\sqrt{2r+1}$, so $2r+1=s^2$ for some odd integer $s$. Now,
$T_{\Lambda_1,\Lambda_1}=\i\,\exp[-\pi\i\,{1\over 2s^2}]$ is a
primitive $s^2$-root of 1, but ${ T}(G)$ will have order dividing
$|G|=2s$, and so those matrices cannot be equal. \cqfd
\smallskip

For which groups will the number of primaries be low? Consider the
formula
\be
|\Phi(G)|=\sum_{a\in \T} \; k(C_G(a)), \label{phi}
\ee
where $k(H)$ is the {class number} of $H$, i.e.\ the number of conjugacy
classes in, or irreducible representations of, $H$. Note that the smaller
$k(G)$ is, the fewer summands there will be in (\ref{phi}), the larger each
conjugacy class $|K_a|$ will tend to be, so the smaller the
centralisers $|C_G(a)|={|G| \over |K_a|}$ will tend to be, and the smaller
the $k(C_G(a))$ in (\ref{phi}) will tend to be. Thus, we should expect
$|\Phi|$ to grow with $k(G)$. The groups with class number less than 13 are
classified \cite{VL}. This allows all $G$ with at most 77 primaries to be
listed. We make this argument precise in the proof of Thm. 2 below.

When $G$ is abelian, it has $|G|^2$ primaries -- this is the extreme case.
The number of primaries for the even dihedral groups ${\frak D}_{2n}$, and
the  quaternion group ${\frak Q}_{2n}$, are both
$2n^2+14$, which grows like $|G|^2/8$.
The odd dihedral groups ${\frak D}_{m}$, $m$ odd, have ${m^2 + 7 \over 2}$
primaries (see below). By comparison (using the data given in Ch.20 of
\cite{BZ}), SL$_2({\Bbb F}_q)$ has $q^2+8q+9$ primaries for $q$ any power
of an odd prime,
and $q^2+q+2$ primaries for $q$ any power of 2 -- in either case, that
number grows like $|G|^{2/3}$. Our computations so far suggest the
following rule of thumb: the more abelian the group is, the messier it
behaves (i.e. the more its primaries, the more its physical invariants, etc),
while the closer the group is to being non-abelian simple, the better
behaved  it will be.

 From this point of view, an interesting measure of how complicated
a group is relative to its size, is the ratio
\be
{\cal N}(G):={{\rm log}\,|\Phi(G)| \over {\rm log}\,|G|}.
\ee
It ranges from 0 to 2, with 2 achieved iff $G$ is abelian. How
low can ${\cal N}(G)$ be ? Some small values are ${\cal N}({\frak
A}_5)\approx .75$, ${\cal N}({\frak A}_7)\approx .55$ and ${\cal
N}(M_{11})\approx .49$. Sporadic simple groups like the Monster
should have ${\cal N}$ very small.

\smallskip\noindent{{\bf Theorem 2.}} There are precisely 33 groups $G$
with at most 50 primaries:

\begin{itemize}

\item the abelian groups
$G={\c}_1,\c_2,\c_3,\c_4,\c_2\times\c_2,\c_5,\c_6,
\c_7$, with precisely $|G|^2$ primaries;

\item the symmetric and alternating groups ${\frak S}_3,{\frak A}_4,
{\frak S}_4,{\frak A}_5,{\frak S}_5,{\frak A}_6$
with $(|G|,|\Phi|)=(6,8),$ (12,14),(24,21),(60,22),(120,39),(360,44);

\item the (semi)dihedral and quaternion groups ${\frak D}_5,{\frak D}_4,
{\frak Q}_4,{\frak D}_7,{\frak D}_9,
{\frak D}_8,{\frak Q}_8,S{\frak D}_8$ with
$(|G|,$ $|\Phi|)=(10,16),(8,22),
(8,22),(14,28),(18,43),(16,46),(16,46),(16,46)$;

\item the Frobenius groups $\c_5\times_f \c_4$, $\c_7\times_f \c_3$,
$\c_3^2\times_f \c_2$, $\c_3^2\times_f \c_4$, $\c_3^2\times_f {\frak
Q}_4$,
$\c_7\times_f
\c_6$, $\c_{11}\times_f \c_5$ with
$(|G|,|\Phi|)=(20,22),(21,32),(18,44),
(36,36),(72,32),(42,44),(55,49)$;

\item the remaining groups $\c_2\times {\frak S}_3$, $DC_3$,
PSL$_2({\Bbb F}_7)$, and SL$_2({\Bbb F}_3)$, with
$(|G|,|\Phi|) = (12,44)$, (12,32), (168,35), (24,42).

\end{itemize}

The semidihedral group $S{\frak D}_{4m}$ is defined by the
presentation
\be
 S{\frak D}_{4n} = \langle r,s \;|\; r^{4n} = s^2 = e, \, srsr = r^{2n}
\rangle.\label{SD}
\ee
 and has order $8m$. The other groups are
defined in \cite{VL}. An interesting consequence of Theorem 2, as mentioned
earlier, is that all groups with $|\Phi(G)|\le 50$ are uniquely
determined by their matrix $S$.

A useful quantity for proving Thm.\ 2 is $h(n)$, the minimum possible class
number $k(H)$ for  $H$ with order $n$ and non-trivial centre. Knowing
$h(n)$ gives a lower bound for $|\Phi(G)|$ once we know the orders
$|C_G(a)|$ of its centralisers.
\cite{VL} can be used for the smaller $n$, basic results on
the classification of finite groups,
as well as the congruence $k(H)\equiv |H|$ (mod 16) when $|H|$ is odd,
give us other $n$. What we find is e.g.\ $h(n)=n$
for all $n\le 27$, except for $n=8,12,16,18,20,24$ (with $h(n) =
5,6,7,9,8,7$, resp). Also useful is the largest value $\ell(k)$ of $|H|$,
for $H$ with
non-trivial centres and class number $k(H)\le k$. For instance for
$k=1,2,\ldots,8$ we get $\ell=1,2,3,4, 8,12,24,48$.

Consider the smallest $|\Phi|$ can be if $k(G)\ge 13$. Note that
at most one $a\in \T$ can have a centraliser $C_G(a)$ with class number
2 --- otherwise if there were two then together exactly
$|G|/\ell(2)+|G|/\ell(2)>|G|-1$ elements would be in the
(disjoint) conjugacy classes $K_a\cup K_b\subset G-e$. More generally,
at most $\ell(k)-1$  $a$'s could have $k(C_G(a))\le k$.
Thus (\ref{phi}) will be bounded
below by
$13+2+3+4+4\cdot 5+4\cdot 6+7=73$. Tightening the argument (e.g.\ ${1\over 2}
+{1\over 3}+{1\over 4}$ is too big) gives $|\Phi|\ge 78$. Thus using
the tables of \cite{VL} we would be able to find all groups with at most
77 primaries. Similarly, to do
$|\Phi|\le 50$ it is enough to consider $k(G)\le 9$. In \cite{VL} are
also given the orders of the centralisers $C_G(a)$. It is
now straightforward to get the Theorem. \cqfd


\section{Examples}

In this section we give a number of explicit examples of untwisted modular
data. We also identify their physical invariants in some cases,
leading to a perplexing situation we will discuss more fully next
section.

Incidentally, a simple construction is direct product:
the modular data for the direct product $G\times H$ is easily obtained from
that of $G$ and $H$. For example, $\Phi({G\times H}) =
\Phi(G)\times\Phi(H)$, ${ S}({G\times H})$ is the Kronecker matrix
product ${ S}(G)\otimes { S}(H)$, etc. Of course, semi-direct
product in general will be much more difficult to work out.

\subsection{Abelian groups}

Abelian $G$ (untwisted) is trivial to work out, but also very uninteresting.
Write $G$ in the following canonical way:
$G\cong\c_{d_1}\times\c_{d_2}\times\cdots\times\c_{d_s}$ where
$d_1|d_2|\cdots|d_s$. For convenience, define a bilinear form on
$\Z^s$ by $\langle m,n\rangle=\sum_i{m_in_i\over d_i}$.
We can identify $\Phi$ here with the $2s$-tuple
$(m,n)\in\Z^s\times\Z^s$, where $0\le m_i\le d_i-1$ and $0\le n_i\le d_i-1$:
in particular, $m$ corresponds to the group element
$(m_1,m_2,\ldots,m_s)\in {\c}_{d_1}\times\cdots\times{\c
}_{d_s}$, and $n$ corresponds to the character $\varphi_n$ of $G$
defined by $\varphi_n(m)=\exp[2\pi\i\langle m,n\rangle]$.
The matrices $S$ and $T$ are given by
\be
S_{(m,n),(m',n')}={1\over |G|}\,\exp[-2\pi\i(\langle m',n\rangle+\langle
m,n'\rangle)]
,\qquad T_{(m,n),(m,n)} =\exp[2\pi\i\langle m,n\rangle].
\ee

All $(m,n)\in\Phi$  are simple currents, with composition given by
pairwise addition. Charge conjugation takes $(m,n)$ to
$(-m,-n)$, and more generally the $\ell$th Galois automorphism (for
$\ell\in \Z_{d_s}^{\times}$) sends $(m,n)$ to $(\ell m,\ell n)$.

$G$ will have many physical invariants, but they  can all be most
elegantly interpreted
using lattices as was explained in \cite{Ga}
(the standard reference for lattice theory is \cite{CS}).
In particular let $\Lambda$ be the $4s$-dimensional integral
indefinite lattice, given by the orthogonal direct sum
$\Lambda=\sqrt{d_1}II_{2,2}
\oplus\sqrt{d_2}II_{2,2}
\oplus\cdots\oplus\sqrt{d_s}II_{2,2}$, where $II_{2,2}=II_{1,1}\oplus
II_{1,1}$ is the unique 4-dimensional even self-dual indefinite
lattice. Then there is a natural one-to-one bijection
between the physical invariants of $G$ and the even self-dual `gluings' of
$\Lambda$, i.e.\ the even self-dual $4s$-dimensional lattices
containing $\Lambda$.
For instance, for $G=\c_p$ ($p$ prime), one finds that there are
precisely  6 (if $p=2$) or 8 (if $p>2$)  physical invariants. Two of these are
\be
(\sum_{i=0}^{p-1} \; {\rm ch}_{i0})\,(\sum_{j=0}^{p-1}{\rm ch}_{0j}^* )
\qquad {\rm
and} \qquad \sum_{i,j=0}^{p-1} \; {\rm ch}_{ij}\,{\rm ch}_{-j,-i}^*\,,
\ee
using obvious notation.

The non-abelian groups are much more interesting, and we turn to them in the
next subsection.
The number of non-abelian groups of order $n \le 50$ are 1 (for
$n=6,10,14,21,22,26,$
$34,38,39,46$), 2 ($n=8,27,28,44$), 3 ($n=12,18,20,30,50$),
5 (for $n=42$), 9 (for $n=16$), 10 (for $n=36$), 11 (for $n=40$), 12 (for
$n=24$), 44 (for $n=32$), 47 (for $n=48$), and 0 otherwise.

\bigskip

\subsection{Some infinite series}

It is not hard to work out $S$ and
$T$ for the infinite series ${\frak D}_n$ (dihedral) and ${\frak Q}_{2n}$
(quaternion):
\bea
&& {\frak D}_n = \langle r,s \;|\; r^n = s^2 = e, \, rsrs=e \rangle, \\
&& {\frak Q}_{2n} =
\langle r,s \;|\; r^{2n}=e, \, s^2=r^n, \, rsrs^{-1} = e \rangle,
\eea
The character tables of ${\frak D}_{4n}$ and ${\frak Q}_{4n}$ are identical.

Consider first the even dihedral groups ${\frak D}_{2n}$, of order
$4n$. ${\frak D}_{2n}$ has $n+3$ conjugacy classes, with representatives
$\T=\{e,\, r^n,\, r^k \;(1 \leq k \leq n-1),\, s,\, sr\}$. It has an
equal number of irreducible representations; 4 are one-dimensional and
$n-1$ are
two-dimensional. The characters and centralisers of the various classes
are indicated in the following table.

\medskip
\renewcommand{\arraystretch}{1.5}
\begin{center}
\begin{tabular}[t]{c|ccccc}
 & $e$ & $r^n$ & $r^k$ & $s$ & $sr$ \\
\hline
$\psi_0$ & 1 & 1 & 1 & 1 & 1\\
$\psi_1$ & 1 & 1 & 1 & $-1$ & $-1$ \\
$\psi_2$ & 1 & $(-1)^n$ & $(-1)^k$ & 1 & $-1$ \\
$\psi_3$ & 1 & $(-1)^n$ & $(-1)^k$ & $-1$ & 1 \\
$\chi_i$ & 2 & $2(-1)^i$ & $2\cos{\pi ik \over n}$ & 0 & 0\\
\hline
$C_G(g)$ & ${\frak D}_{2n}$ & ${\frak D}_{2n}$ & $\Z_{2n}$ &
$\Z_2 \times \Z_2$ & $\Z_2 \times \Z_2$
\end{tabular}
\end{center}
\medskip

It follows that the number of primaries is equal to $|\Phi| = 2n^2 + 14$.
Denote by $\psi_i$ ($0\le i<2n$) the characters of $\langle
r\rangle\cong \Z_{2n}$, given by $\psi_i(r^k)=\xi_{2n}^{i k}$.
Likewise, denote by $\phi_{ab}$ and $\varphi_{ab}$ ($0\le a,b\le 1$)
the characters of $\langle s,r^n\rangle\cong \Z_2\times\Z_2$ and
$\langle rs,r^n\rangle\cong \Z_2\times\Z_2$, respectively, where
$\phi_{ab}(s^kr^{n\ell})=\varphi_{ab}((rs)^kr^{n\ell})=(-1)^{ak+b\ell}$.

The values of the diagonal entries of $T$ follow directly from (\ref{uT}).
Eq. (\ref{cen}) easily computes any $S$ entry involving $e$ or $r^n$. The
remaining non-zero $S$ entries are
\bea
&&S_{(r^k,\psi_i),(r^\ell,\psi_j)}={1\over n} \cos(\pi{\ell i+kj\over n}),\\
&&S_{(s,\phi_{ab}),(s,\phi_{cd})}=S_{(rs,\varphi_{ab}),(rs,\varphi_{cd})}=
{1\over 4}\cdot\left\{\matrix{(-1)^{a+c}+(-1)^{a+b+c+d}&{\rm if}\ n\ {\rm
even,}\cr
(-1)^{a+c}&{\rm if}\ n\ {\rm odd,}}\right.
\eea
and in addition for $n$ odd
$S_{(s,\phi_{ab}),(sr,\varphi_{cd})}={1\over 4}(-1)^{a+b+c+d}$.

The odd dihedral groups ${\frak D}_{2n+1}$ can be worked out in the same
way. ${\frak D}_{2n+1}$ has $n+2$ conjugacy classes, with representatives
$\T=\{e,\, r^k \; (1 \leq k \leq n),\, s\}$. It has two one-dimensional
representations, and $n$ two-dimensional representations. The characters
and centralisers of the various classes are reproduced in the following
table.

\medskip
\begin{center}
\begin{tabular}[t]{c|ccc}
 & $e$ & $r^k$ & $s$ \\
\hline
$\psi_0$ & 1 & 1 & 1 \\
$\psi_1$ & 1 & 1 & $-1$ \\
$\chi_i$ & 2 & $2\cos{2\pi ik \over 2n+1}$ & 0\\
\hline
$C_G(g)$ & ${\frak D}_{2n+1}$ & $\Z_{2n+1}$ & $\Z_2$
\end{tabular}
\end{center}
\medskip

One finds that the number of primary fields is $|\Phi| = 2n^2 + 2n + 4$.
Write $\psi_i$ for the characters of $\langle r\rangle\cong\Z_{2n+1}$
as before, and $\varphi_i$ for the obvious two characters of $\langle
s\rangle\cong \Z_2$.
The calculation of $S$ and $T$ proceeds like for the even dihedral groups:
\bea
&& S_{(r^k,\psi_i),(r^\ell,\psi_j)}={2\over 2n+1}\cos(2\pi{kj+\ell i\over
2n+1}),\\
&& S_{(s,\varphi_i),(s,\varphi_j)}={1\over 2}(-1)^{i+j}.
\eea

The semidihedral groups $S{\frak D}_{4m}$ (2.27) have the same matrix $T$ as
${\frak D}_{4m}$, but their matrix $S$ is always complex.

Next turn to the quaternions ${\frak Q}_{2n}$.
It has $n+3$ conjugacy classes, with representatives
$\T=\{e,\, r^n,\, r^k \;(1 \leq k \leq n-1),\, s,\, sr\}$. It has 4
 one-dimensional and $n-1$
two-dimensional representations. The characters and centralisers of the
various
classes are indicated in the following table (put $\iota=1$ for $n$
even, and $\iota=\i$ for $n$ odd).

\medskip
\renewcommand{\arraystretch}{1.5}
\begin{center}
\begin{tabular}[t]{c|ccccc}
 & $e$ & $r^n$ & $r^k$ & $s$ & $sr$ \\
\hline
$\psi_0$ & 1 & 1 & 1 & 1 & 1\\
$\psi_1$ & 1 & 1 & 1 & $-1$ & $-1$ \\
$\psi_2$ & 1 & $(-1)^n$ & $(-1)^k$ & $\iota$ & $-\iota$ \\
$\psi_3$ & 1 & $(-1)^n$ & $(-1)^k$ & $-\iota$ & $\iota$ \\
$\chi_i$ & 2 & $2(-1)^i$ & $2\cos{\pi ik \over n}$ & 0 & 0\\
\hline
$C_G(g)$ & ${\frak Q}_{2n}$ & ${\frak Q}_{2n}$ & $\Z_{2n}$ &
$\Z_4$ & $\Z_4$
\end{tabular}
\end{center}
\medskip

The number of primaries is equal to $|\Phi| = 2n^2 + 14$.
Denote by $\psi_i$  the $2n$ characters of $\langle
r\rangle\cong \Z_{2n}$, and  by $\phi_{a}$ and $\varphi_{a}$
the 4 characters of $\langle s\rangle\cong \Z_4$ and
$\langle rs\rangle\cong \Z_4$, respectively.
The nonzero $S$ entries not involving  $e$ or $r^n$ are
\bea
&& S_{(r^k,\psi_i),(r^\ell,\psi_j)}={1\over n} \cos(\pi{\ell i+kj\over
  n}),\\
&& S_{(s,\phi_{a}),(s,\phi_{b})}=S_{(rs,\varphi_{a}),(rs,\varphi_{b})}=
{1\over 4}\cdot\left\{\matrix{\i^{a-b}&{\rm if}\
    n\ {\rm odd,}\cr 2\cos(\pi{a+b\over 2})&{\rm if}\ n\ {\rm even,}}\right.
\eea
and in addition for $n$ odd $S_{(s,\phi_{a}),(sr,\varphi_{b})} = {1\over
4}\i^{a+b}$.

For our final example in this subsection, we will consider the
series of non-abelian simple groups, SL$_2({\Bbb F}_q)$ for $q=2^n$. Note
that SL$_2({\Bbb F}_2)\cong {\frak S}_3$ and SL$_2({\Bbb F}_4)\cong {\frak
A}_5$.

The order of SL$_2({\Bbb F}_q)$ is $q(q^2-1)$. There are $q+1$ conjugacy
classes whose representatives can be chosen in
\be
\T=\Big\{e,\, \iota = \pmatrix{1 & 0 \cr 1 & 1},\,
\alpha^a = \pmatrix{s^a & 0 \cr 0 & s^{-a}}, \, \beta^b \;:\;
1 \le a \le {\textstyle {q-2 \over 2}} \;{\rm and} \; 1 \le b \le
{\textstyle {q \over 2}} \Big\},
\ee
where $s$ is any generator of the cyclic multiplicative group ${\Bbb
F}_q^*$, and where $\beta$ is an
element of order $q+1$ (its exact form is not important). The
characters (the labels $i$ and $j$ run from 1 to ${q \over 2}-1$ and $q
\over 2$ respectively) and centralisers are given below, from which one
finds the number of primaries equals $|\Phi| = q^2 + q + 2$, as mentioned
before.

\medskip
\begin{center}
\begin{tabular}[t]{c|cccc}
 & $e$ & $\iota$ & $\alpha^a$ & $\beta^b$ \\
\hline
$\psi_0$ & 1 & 1 & 1 & 1\\
$\psi_1$ & $q$ & 0 & 1 & $-1$ \\
$\chi_i$ & $q+1$ & 1 & $2\cos{2\pi ia \over q-1}$ & 0\\
$\theta_j$ & $q-1$ & $-1$ & 0 & $-2\cos{2\pi jb \over q+1}$ \\
\hline
$C_G(g)$ & SL$_2({\Bbb F}_q)$ & $\Z_2^n$ & $\Z_{q-1}$ & $\Z_{q+1}$
\end{tabular}
\end{center}
\medskip

Because  $2,q-1,q+1$ are pairwise coprime, Proposition 1(a) says that the
only potentially nonvanishing $S_{(a,\chi),(b,\phi)}$ entries have
$a=e$ or $b=e$, or $a=b=\iota$, or $a,b$ are both powers of $\alpha$,
or $a,b$ are both powers of $\beta$. The relevant sets $G(g,h)$ here are
$G(\iota,\iota)=\cup_{ a=0}^{q-2} \left({1 \atop *} \; {0 \atop
1}\right)\alpha^a$, $G(\alpha^a,\alpha^b)= \langle a\rangle\cup\langle
a\rangle\left({0 \atop 1} \; {1 \atop 0}\right)$, and
$G(\beta^a,\beta^b)=\langle \beta\rangle\cup \langle \beta\rangle\gamma$
where $\gamma^{-1}\beta\gamma=\beta^{-1}$.

Explicitly writing down its matrix $S$ would require the evaluation of
some interesting character sums, something we have not yet done.

\subsection{The physical invariants for ${\frak S}_3$}

The non-abelian group of smallest order is ${\frak S}_3$. Its character
table is

\medskip
\begin{center}
\begin{tabular}[t]{c|ccc}
 & $e$ & (123) & (12) \\
\hline
$\psi_0$ & 1 & 1 & 1 \\
$\psi_1$ & 1 & 1 & $-1$ \\
$\psi_2$ & 2 & $-1$ & 0\\
\hline
$C_G(g)$ & ${\frak S}_3$ & $\Z_3$ & $\Z_2$
\end{tabular}
\end{center}
\medskip

The modular data for ${\frak S}_3$ will have 8 primary fields: $(e,\psi_i)$
for $i=0,1,2$; $((123),\varphi_k)$, $k=0,1,2$, for the 3 characters
$a\mapsto \xi_3^{ak}$ of $\c_3$; and $((12),\varphi'_k)$, $k=0,1$,
for the 2 characters $b\mapsto (-1)^{bk}$ of $\c_2$. For convenience
label these primaries $0,1,\ldots, 7$.

Since ${\frak S}_3\cong {\frak D}_3$, we can read
${ S}({\frak S}_3)$ and ${ T}({\frak S}_3)$ off from the
previous subsection:
\be
S={1\over 6}\,\left(\matrix{1&1&2&2&2&2&3&3\cr 1&1&2&2&2&2
&-3&-3\cr 2&2&4&-2&-2&-2&0&0\cr
2&2&-2&4&-2&-2&0&0\cr
2&2&-2&-2&-2&4&0&0\cr
2&2&-2&-2&4&-2&0&0\cr
3&-3&0&0&0&0&3&-3\cr 3&-3&0&0&0&0&-3&3\cr}\right),
\ee
\be
T={\rm diag}(1,1,1, 1,\xi_3,\xi_3^2,1,-1).
\ee

There is one non-trivial simple current $(e,\psi_1)$ (namely primary
\#1), identifiable by the 1 in the corresponding entry of the 0th row
of $S$.
Since all entries of $S$ are rational, the charge conjugation and the
other Galois permutations $\si_\ell$ are trivial. Incidentally, groups $G$ for
which ${ S}(G)$ is rational are rare; that property requires that
the exponent of $G$ divides 24 (e.g.\ the exponent of ${\frak S}_3$ is 6).
To see that, apply Prop.1(b) to the equation
$(T_{a,a})^{\ell^2}=T_{\si_\ell a,\si_\ell a}$.

There are precisely 32 physical invariants for ${\frak S}_3$. Write ch$_i$
for the CFT
character corresponding to the $i$th primary. The automorphism
invariants are $M=I$, and the one (call it $M'$) which interchanges
$2\leftrightarrow 3$ and fixes everything else. Extending by the simple
current gives us 3 invariants:
\be
|{\rm ch}_0+{\rm ch}_1|^2+2|{\rm ch}_2|^2+2|{\rm ch}_3|^2+2|{\rm
  ch}_4|^2+2|{\rm ch}_5|^2+k({\rm ch}_2{\rm ch}_3^*+{\rm ch}_3{\rm
  ch}_2^*-|{\rm ch}_2|^2-|{\rm ch}_3|^2),
\ee
for $k=0,1,2$. Write $s_1:={\rm ch}_0+{\rm ch}_1+{\rm ch}_2+{\rm
  ch}_3$,  $s_2:={\rm ch}_0+{\rm ch}_1+2{\rm ch}_2$,
$s_3:={\rm ch}_0+{\rm ch}_1+2{\rm ch}_3$, $s_4:={\rm ch}_0+{\rm
  ch}_2+ {\rm ch}_6$, and $s_5:={\rm ch}_0+{\rm ch}_3+{\rm ch}_6$; then
$s_is_j^*$ is a physical invariant. The final ones are
\be
|{\rm ch}_0+{\rm ch}_1+{\rm ch}_2+{\rm ch}_3|^2+k({\rm ch}_2{\rm
  ch}_3^*+ {\rm ch}_3{\rm ch}_2^*-|{\rm ch}_2|^2-|{\rm ch}_3|^2),
\ee
for $k=\pm 1$.

In the next section we manage to identify general constructions
yielding  most of these
physical invariants. Two however remain unexplained: the automorphism
invariant $M'$, and the modular invariance of the sum $s_1$.

Note that in the basis defined by ($\xi = \xi_3 = {\rm exp}[2 \pi\i /3] $)
\bea
(e_0, \ldots , e_7) &\!\!=& \!\!
({\rm ch}_0  - {\rm ch}_1 \,,\,
{\rm ch}_0  + {\rm ch}_1  + 2{\rm ch}_2 \,,\,
{\rm ch}_0  + {\rm ch}_1  -  {\rm ch}_2  \,,\,
{\rm ch}_3  + {\rm ch}_4  +  {\rm ch}_5  \,,\, \nonumber\\
&& {\rm ch}_3  + \xi   {\rm ch}_4  + \xi^2 {\rm ch}_5  \,,\,
{\rm ch}_3  + \xi^2 {\rm ch}_4  + \xi   {\rm ch}_5  \,,\,
{\rm ch}_6  + {\rm ch}_7  \,,\,
{\rm ch}_6  - {\rm ch}_7),
\eea
the matrices $S$ and $T$ become permutation matrices (both non diagonal),
corresponding respectively to the permutations (06)(23)(45) and (345)(67) of
${\frak S}_8$ (written in terms of cycles). From this, it is not difficult
to compute the dimension (over $\C$) of the commutant of $S$ and $T$, which
turns out to be 11. Hence the 32 physical invariants are not linearly
independent
(e.g. $s_2 + s_3 = 2s_1$ and $2s_4 - 2s_5 = s_2 - s_3$).

Incidentally, it is a consequence of \cite{DVVV} that for any $G$, such a
basis can always be found in which $S$ and $T$ are permutation
matrices. In particular, the characters ch$_{(a,\chi)}$, for
$(a,\chi)\in\Phi$, can be thought of as a function on $G\times G$
taking any pair $(xax^{-1},xbx^{-1})$ to $\chi(b)$ when $a$ and $b$
commute, and sending all other pairs in $G\times  G$ to 0. They span a
space called $C^0(G_{comm})$ in \cite{KSSB}. Now choose any
commuting pair $(a,b)\in G\times G$, and define the function
$f_{(a,b)}$ to be identically 1 on the set $(xax^{-1},xbx^{-1})$ and 0
elsewhere. Then these functions form another basis for
$C^0(G_{comm})$. For this choice of basis, our representation of
SL$_2(\Z)$ becomes manifestly a permutation representation:
$\left(\matrix{a&b\cr c&d}\right)\in{\rm SL}_2(\Z)$ acts on the right by
sending $f_{(g,h)}$ to $f_{(g^ah^c,g^bh^d)}$. For instance, we recover
$T$ by noting that 
\be{\rm  ch}_{(a,\chi)}(a,ab)=\chi(ab)={\chi(a)\over \chi(e)}\,
\chi(b)=T_{(a,\chi),(a,\chi)}{\rm ch}_{(a,\chi)}\ .\ee
(See equation (5.12) of
\cite{KSSB} for the corresponding matrix $S$ calculation.)
It is a very
special property of the (untwisted) finite group modular data
that this SL$_2(\Z)$ representation is actually a permutation
representation. For
general modular data, usually the matrix $T$ won't even be conjugate to a
permutation matrix. Two examples are the simplest affine data, i.e.\
$A_1^{(1)}$ at level 1, and the simplest twisted group data, $G=\Z_2$ with
twist $\alpha_1$ (see \S6.1), whose matrices $T$ are respectively
\be
{\rm diag}\left(\exp[-\pi\i/ 12],\exp[5\pi\i/12]\right)\qquad
{\rm and}\qquad {\rm diag}\left(1,1,\i,-\i\right)\ .
\ee
The first cannot be similar to a permutation matrix because it does not
have 1 as an eigenvalue, while the second cannot because an order 4
permutation matrix has $-1$ as an eigenvalue.

Incidentally, the analogue of this permutation representation for the
twisted finite group data is known \cite{B3}, and will be discussed in \S5.3.

A total of 32 physical invariants for such a small number
of primaries is completely unprecedented from the more familiar WZW
situation. The orthogonal algebras at level 2 are the worst behaved affine
cases, but e.g. $D_{16}^{(1)}$ at level 2 has 23 primaries but only 22
physical invariants. But ${\frak S}_3$, being a dihedral group, is nearly
abelian so should be especially bad in this respect.

It seems that the twisted data is better behaved in this respect than the
untwisted.
For example, we find, combining the discussion of \S6.3 with the results of
\cite{G}, that the twisted modular data for ${\frak S}_3$ has only 9 physical
invariants.


\section{Making sense of all the physical invariants}

We learned in the previous section that there is a surprising number
of physical invariants associated to finite group modular data. In this
section we try to tame the zoo!

The conventional wisdom in RCFT is that all physical invariants are
constructed in two ways: as extensions of chiral algebras, and as
automorphisms of those chiral algebras (called automorphism invariants
in the unextended case).
In the affine case, the physical invariants are almost always `obvious'
in hindsight, as symmetries of the appropriate Dynkin diagram (e.g.\ simple
currents and charge conjugation) are directly responsible for almost all
WZW physical invariants.
It would be nice to find the analogous statement for the finite group
modular data.

A rich supply of physical invariants always comes from simple currents
\cite{Be,KS}. As these are well-understood, we won't say any more than we
did in \S2.

\subsection{Automorphism invariants}

The outer automorphism group Out($G$) provides a systematic way of constructing
automorphism invariants. Choose any $\pi\in{\rm Out}(G)$. Then $\pi$ induces
an invertible homomorphism $C_G(a)\rightarrow C_G(\pi a)$ between
centralisers. Define $\chi^\pi$ by the formula
$\chi^\pi(g)=\chi(\pi^{-1}g)$ --- if $\chi\in{\rm Irr}(C_G(a))$ then $\chi^\pi
\in{\rm Irr}(C_G(\pi a))$. Consider the permutation of $\Phi$ defined
by $(a,\chi)\mapsto(\pi a,\chi^\pi)$. Then it commutes with $S$ and $T$
and hence defines an automorphism invariant.

Many (but not all) outer
automorphisms can be interpreted in the following way:
Any time a given group $G$ is a normal subgroup of another group
$\widehat{G}$, then any element $\hat{g}\in\widehat{G}$ defines an
automorphism of $G$ by conjugation: $g\mapsto\hat{g}g\hat{g}^{-1}$.

Examples of these groups are Out($\Z_n)\cong\Z^\times_n$;
Out$({\frak D}_{2k+1})\cong\Z_{2k+1}^\times/\{\pm 1\}$; for $n\ge 3$,
Out$({\frak
S}_n)=\{1\}$
and Out$({\frak A}_n)\cong{\c}_2$, except Out$({\frak S}_6)\cong {\c}_2$
and  Out$({\frak A}_6)\cong {\c}_2\times{\c}_2$.

Another systematic source are the Galois automorphism invariants --- there
is one of these for every $\si_\ell\in{\rm Gal}(\Q(S)/\Q)$ with
$\ell^2\equiv 1$ modulo the exponent $e(G)$ of $G$, and the
permutation is simply given by the Galois permutation, which we may write
$\si_\ell(a,\chi)=(a^\ell,\si_\ell\chi)$. If
there are $s$ distinct primes which divide $e(G)$, then there
will be precisely 
\be\left\{\matrix{2^{s-1}&{\rm if}\ e(G)\equiv 2\ {\rm (mod\ 4)}\cr
2^{s+1}&{\rm if}\  e(G) \equiv 0\ {\rm (mod\ 8)}\cr
2^s&{\rm otherwise}\cr}\right. \ee
such $\ell$,  and hence that number of  Galois
automorphism invariants (though these won't necessarily be distinct, if
$\Q(S)$ is
smaller than $\Q(\xi_{e(G)})$).

A final source of automorphism invariants is {\it discrete torsion} \cite{V}.
This involves the language of cohomology --- see \S5.1 for the
appropriate definitions. Take any 2-cocycle $\beta\in
Z^2(G,U(1))$ for $G$. This $\beta$ is entirely independent of the
2-cocycles $\beta_a$ we discuss in \S5.2.
For each conjugacy class representative $a\in \T$, define $e_a(b):=
\beta(a,b)\beta(b,a)^{-1}$. This will be a 1-dimensional representation of
$C_G(a)$.
Define the permutation of primary fields $(a,\chi)\in \Phi$ sending
$(a,\chi)$ to $(a,e_a\chi)$. It is easy to check that this commutes
with $S$ and $T$ and thus defines an automorphism invariant.

When is this permutation trivial ? Iff each $e_a(b)=1$, for all $b\in C_G(a)$.
In other words, iff each $a\in G$ is $\beta$-regular (see \S5.1 for
the definition).
One thing this means is that cohomologous $\beta$ give the same
discrete torsion. So the possibilities for discrete torsion are given
by the classes in the finite abelian group $H^2(G,U(1))=M(G)$, known
as the Schur multiplier of $G$.
Each class in $M(G)$ will usually but not always give rise to a different
automorphism invariant.
Incidentally, this construction applies to untwisted, CT twisted, or non-CT
twisted,
modular data (see next section for the twisted modular data).

So discrete torsion will always give $|M(G)|$ automorphism invariants, although
these are not necessarily all distinct. They are all distinct
for $\Z_n^2$, $\Z_n^3$, and ${\frak D}_{even}$.

\subsection{Chiral extensions}

Making sense of the large number of physical invariants also requires finding
the ``generic'' chiral extensions. Many will come from simple currents.
It is tempting to suspect that another  rich source could be from
certain  normal subgroups $N\triangleleft G$.

In particular, Clifford theory (which is concerned with the theory of
induced and restricted characters) can be used to explicitly relate the
modular data
of a group $G$ to that of  its normal subgroups $H$. The  formulas we have
 obtained however are complicated enough that at this point we have been
unable to
 determine explicit relations between the physical invariants of $G$
 and $H$. Nevertheless, the situation is sufficiently analogous to
 that of conformal embeddings \cite{kw} in affine algebras that we
 expect this to likewise be a rich source of chiral
 extensions. Incidentally, there are also relations between the modular
 data of $G$ and $G/N$.

Nevertheless we have found some interesting generic chiral extensions.
Choose any central elements $z,z'\in Z(G)$  and any degree-1
characters $\varphi,\psi\in{\rm Irr}(G)$. Define the combinations
\bea
s_{z,z'}:=&\sum_{\chi\in {\rm Irr}(G)} \chi(z')\,{\rm ch}_{(z,\chi)}\\
s'_{\varphi,\psi}:=&\sum_{g\in \T}\psi(g)\,{\rm ch}_{(g,\varphi)}
\eea
Then it is easy to verify, using (\ref{rest}), Frobenius reciprocity,
and the orthogonality relations for characters,  that
$s_{z,z'}(-1/\tau)=s_{z',z^{-1}}(\tau)$ and $s'_{\varphi,\psi}(-1/\tau)=
s'_{\psi,\varphi^*}(\tau)$, while 
$s_{z,z'}(\tau+1)=s_{z,zz'}(\tau)$ and $s'_{\varphi,\psi}(\tau+1)=
s'_{\varphi,\varphi\psi}(\tau)$.
To see this, consider the coefficient of ch$_{(a,\chi)}(\tau)$ in 
$s'_{\varphi,\psi}(-1/\tau)$: it will be
\be
\sum_{g\in \T}\psi(g)\,{1\over|G|}\,\varphi(a)^*\chi_{C_G(a)}^G(g)^*
\ee
which is $\varphi(a)^*$ times the coefficient of $\psi$ in
$\chi_{C_G(a)}^G$, i.e.\ $\varphi(a)^*$ times the coefficient of $\chi$
in $\psi|_{C_G(a)}$, i.e.\ $\varphi(a)^*\delta_{\psi,\chi}$.

Choose any subgroups $A\le Z(G)$, $B\le G/G'$, and define the sums
\bea
s(A)&=&{1\over |A|}\sum_{z,z'\in A}s_{z,z'}=\sum_{z\in
  A}\sum_{\chi:{\rm ker}(\chi)\ge A}\chi(e)\,{\rm ch}_{z,\chi}\\
s'(B)&=&{1\over |B|} \sum_{\varphi,\psi\in B}s'_{\varphi,\psi}=
\sum_{\psi\in B}\sum_{g\in {\rm ker}(B)\cap \T}{\rm ch}_{g,\psi}
\eea
By ker$(\chi)$ we mean all $g\in G$ for which $\chi(g)=\chi(e)$, and
by ker$(B)$ we mean the intersection of all ker$(\psi)$, where
$\psi\in B$ is identified with a degree-1 character of $G$.
Note that the sums $s(A),s'(B)$ are all invariant under $S$ and $T$.
Thus {\it for any $G,A,A',B,B'$} we get the remarkable physical invariants
\be
s(A)^*\,s(A'), \quad s(A)^*s'(B), \quad s'(B)^*s(A), \quad s'(B)^*\,s'(B').
\ee
In the case of ${\frak S}_3$ which we worked out in \S3.3,
$s(e)= s_2$, $s'(\psi_0)=s_5$, and $s'(\psi_0,\psi_1)=s_3$. This alone
accounts for 9 of the 32 physical invariants for ${\frak S}_3$.

Incidentally, taking $A$ and $B$ to be the trivial groups completes
the proof of Proposition 1(h): the fifth
physical invariant of course is the diagonal sum. For the second claim
there, there is a natural degree-preserving bijection (see \S3.5 of
\cite{BZ}) between the $\chi\in{\rm Irr}(G)$ with ker$(\chi)\ge Z(G)$,
and Irr$(G/Z(G))$, so if $G/Z(G)$ is nonabelian there will always be
degree $>1$ characters appearing in each $s(A)$ and so they will be
different from any $s'(B)$.


\section{Twisting, Group Cohomology, and Projective Representations}

As advertised in the introduction, one way to generalize the group data
described in Section 2 is by introducing some ``twisting''. This
twisting has a cohomological origin, as in the theory of affine algebras
\cite{K}, where the infinitely many possible twists are labelled by the
level $k$, an integer. In constrast, the twisting of the finite group
modular data offers but a finite number of possibilities.

The twisting of this modular data was first described in the most generality
by \cite{DW,DPR,AC} (see also \cite{DVVV}), although to our knowledge
explicit expressions for the modular matrices $S$ and $T$ in the most
general case have not appeared until now (though the most general fusions
appear in (6.44) of  \cite{DW}).
We will recall here their construction, as concretely as possible, and then
we will compute explicit examples, making contact with known structures
(most notably affine algebras).

\subsection{Cohomological preliminaries}

As a twisting of the finite group modular data is effected by elements of
cohomology groups $H^i(G,U(1)) \cong H^i(G,\C^\times) \cong H^{i+1}(G,\Z)$, we
will first review the relevant properties and give some examples. For our
purposes it is not very important to know how these are defined. We refer the
reader to standard textbooks (like \cite{Br}) for a more complete treatment.
For projective representations, see \cite{K2}.

For all $i>0$, $H^i(G,U(1))$ is a finite abelian group, which we will
usually write additively,  obeying
$|G| H^i(G,U(1))=0$ (i.e.\ all elements have order dividing the order of
$G$). For any $G$, $H^0(G,U(1)) = 0$, while $H^1(G,U(1)) =
G/G'$ is the group of one-dimensional representations of $G$ ($G'$ is the
commutator subgroup). The next group $M(G) := H^2(G,U(1))$, called the
Schur multiplier (or multiplicator), classifies the
(projectively) inequivalent projective representations, and for that
reason, it will play a central role in the whole construction. The only
other group we will be interested in is $H^3(G,U(1))$.

The following results can be useful: the square of the exponent of
$M(G)$ divides $|G|$; if $G$ is a $p$-group (i.e.\ its order is a
power of a prime), then $H^3(G,U(1)) \ne
0$; if $e$ is the exponent of $G$ and $e_i$ is the exponent of
$H^i(G,U(1))$, then $ee_2$, $e_1e_2$ and $e_2e_3$ all divide $|G|$.

Computing cohomology groups, even $M(G)$, is usually difficult.
They are known however for several familiar groups. For a cyclic
group, one has  $M(\Z_n)=0$ and $H^3(\Z_n,U(1))=\Z_n$. For products of
identical cyclic groups,
\be
M(\Z_n^k) = \Z_n^{k(k-1)/2}, \quad H^3(\Z_n^k,U(1)) =
\Z_n^{k(k^2+5)/6}.
\ee

More generally, the Schur multiplier for any abelian group is as follows:
write $G=\Z_{d_1}\times \Z_{d_2}\times\cdots\times\Z_{d_s}$ where
$d_1|d_2|\cdots|d_s$,
then
\be
M(G) = \prod_{j=1}^s \Z_{d_j}^{j-1}.
\ee

Other known groups are (see e.g.\ Ch.6 of \cite{BZ})
\bea
&& M({\frak S}_3) = 0, \quad H^3({\frak S}_3,U(1)) = \Z_6, \quad M({\frak
S}_n) = \Z_2 {\rm \ \ \ for\ } n \ge 4, \\
&& M({\frak A}_n) = \Z_2 {\rm \ \ \ for\ } n \ge 4, \; n \neq 6,7, \quad
M({\frak A}_n) = \Z_6 {\rm \ \ \ for\ } n=6,7,\hskip 1truecm \\
&& M({\frak D}_n) = 0, \quad H^3({\frak D}_n,U(1)) = \Z_{2n}
\qquad {\rm for\ } n {\rm \ odd},\\
&& M({\frak D}_n) = \Z_2, \quad H^3({\frak D}_n,U(1)) = \Z_n \times \Z_2
\times \Z_2  \qquad {\rm for\ } n {\rm \ even},\\
&& M({\frak Q}_n) = 0, \quad H^3({\frak Q}_4,U(1)) = \Z_8, \\
&& M({\rm SL}_n({\Bbb F}_q))=0, \label{sl} \\
&& M({\rm PSL}_n({\Bbb F}_q))=\Z_{{\rm gcd}(q-1,n)}, \label{psl}
\eea
where in (\ref{sl}) and (\ref{psl}), $(n,q) \not \in
\{(2,4),(2,5),(2,7), (2,9),(3,2),(3,3),(3,4),(4,2)\}$.

For odd prime $p$, there are precisely two different non-abelian groups of
order $p^3$: one of these (a split extension of cyclic groups, the $p \ne 2$
analogue of ${\frak D}_4$) has $M=0$ and $H^3=\Z_p \times \Z_p$; the other
has $M=\Z_p \times \Z_p$ and $H^3 = \Z_p \times \Z_p \times \Z_p
\times \Z_p$. (All groups of order $p^2$ are abelian.)

This gives some taste of what $M(G)$ and $H^3(G)$ look like for nice groups.
The Schur multiplier is known for all simple groups. For instance it is
trivial for the Monster.

As mentioned before, the Schur multiplier plays a central role in the
theory of projective representations. Indeed a normalised 2-cocycle $\beta \in
Z^2(G,U(1))$ is a map $G\times G\rightarrow U(1)$ satisfying
$\beta(x,1)=\beta(1,x)=1$ and the
cocycle condition $\beta(x,y)\,\beta(xy,z)=\beta(y,z)\,\beta(x,yz)$ for all
$x,y,z\in G$. For any such $\beta$, one may consider the
projective $\beta$-representations, i.e. the maps $\tilde\rho : G \rightarrow {\rm
GL}(V)$ obeying $\tilde{\rho}(x) \tilde{\rho}(y) = \beta(x,y)
\tilde{\rho}(xy) $. The
cocycle condition corresponds to associativity. If $\beta$ is identically 1,
$\tilde{\rho}$ will be an ordinary representation and will be called
{\it linear}.

If $\beta,\beta'$ are
cohomologous (i.e.\ $\beta^{-1}\beta'$ is a 2-coboundary -- a
{\it 2-coboundary} is any $\beta\in Z^2$ of the form
$\beta=\gamma(x)\gamma(y)\gamma(xy)^{-1}$),  then any
$\beta'$-representation will be projectively equivalent to some
$\beta$-representation, and there will be the same number of
$k$-dimensional $\beta$-representations as $k$-dimensional
$\beta'$-representations, for each $k=1,2,\ldots$. Moreover, a
$\beta$-representation can be one-dimensional only if $\beta$ is a coboundary.

A natural question is to
classify the projective representations belonging to a given cocycle. So let
$r(G,\beta)$ denote the number of linearly inequivalent irreducible
$\beta$-representations of $G$. We know that
$r(G,\beta)$ is a cohomology class invariant.
In order to compute this number, one introduces the notion of a
$\beta$-regular group element: $g \in G$ is $\beta$-regular if $\beta(g,h)
= \beta(h,g)$ for all $h$ in $C_G(g)$. One may check that $g$ is
$\beta$-regular if and only if all its conjugates are, so that a whole
conjugacy class is $\beta$-regular or not. Then
\be
r(G,\beta) = {\rm number\ of\ } \beta{\rm -regular\ conjugacy\
classes\ of\ }G.
\ee
As an immediate consequence, one has the inequality $r(G,\beta) \le  k(G) =
r(G,1)$ for any 2-cocycle $\beta$, with equality if (but not only if)
$\beta$ is a coboundary.

The projective characters $\tilde{\chi}\in\beta$-Irr$(G)$ (defined as usual
by the
trace of  the representation) share many properties with the usual
characters, except
that they are in general not class functions. Schur's lemma still holds,
from which
orthogonality and completeness relations follow: for any
$\tilde{\chi},\tilde{\chi}'\in\beta$-Irr$(G)$ and any $\beta$-regular $a\in G$,
\bea
&& {1 \over {|G|}} \sum_{g \in G}\;\tilde{\chi} (g)^* \, \tilde{\chi}' (g) =
\delta_{\tilde{\chi},\tilde{\chi}'}, \label{orth}\\
&& {{|K_a|} \over|G|} \sum_{\tilde{\chi}\ \beta-{\rm Irr}}\;
\tilde{\chi} (a)^* \, \tilde{\chi}
(b) = \delta_{b \in K_a} \, . \label{compl}
\eea
The dimensions of the irreducible projective representations divide the
order of the group, and their squares sum up to $|G|$. All
$\beta$-characters vanish at non-regular classes.

Examples of projective representation data include the following. Writing
$H^2(\Z_n^2,U(1))=\langle\beta_1\rangle\cong \Z_n$ and
 $H^2(\Z_n^3,U(1)) = \langle\beta_1,\beta_2,\beta_3\rangle\cong \Z_n^3$,
one finds
\be
r(\Z_n^2,\beta_1^x) = [{\rm GCD}(x,n)]^2, \qquad
r(\Z_n^3,\beta_1^x \, \beta_2^y \, \beta_3^z) = n \cdot [{\rm GCD}(x,y,z,n)]^2.
\label{zn3}
\ee
for any $1\le x,y,z\le n$. For the dihedral groups, one has
\be
r({\frak D}_{2n+1},\beta) = n+2, \qquad
r({\frak D}_{2n},\beta) =
\cases{n+3 & if $\beta$ is coboundary,\cr
n & otherwise.}
\ee
These numbers are known also for ${\frak S}_n$ and ${\frak A}_n$.

\subsection{General construction}

The possible twistings are parametrised by the finitely many elements
$\alpha \in H^3(G,U(1))$. The 3-cocycle condition is
$\alpha(g,h,k)\,\alpha(g,hk,\ell)\,\alpha(h,k,\ell)=\alpha(gh,k,\ell)\,
\alpha(g,h,k\ell)$. Different 3-cocycles give rise to different
modular data, with in general different numbers of primary fields (never more
than in the untwisted case), and with matrices $S^{\alpha}$ and
$T^{\alpha}$.

The construction starts from a normalised\footnote{That is, it satisfies
$\alpha(e,h,g) = \alpha(h,e,g) = \alpha(h,g,e) = 1$. This implies that all
2-cocycles are accordingly normalised, $\beta_e(h,g) = \beta_h(e,g) =
\beta_h(g,e)
= 1$ for all $h,g$.} element $\alpha$  of $H^3(G,U(1))$. We can and
will assume for convenience that the values of $\alpha$ are always
roots of 1 -- in fact if $\alpha$ has cohomological order $n$, then
we can require $\alpha$ to take the values of $n$th roots of 1 (proof:
Write $\alpha^n=\delta\beta$; choose any $\gamma=\beta^{-1/n}$, then
$\alpha\gamma$ is cohomologous to $\alpha$ and has the desired
property.) Because $|G|\,H^3(G,U(1))=0$, $n$ will necessarily divide
$|G|$.

For all $a, g, h \in G$, we define auxiliary quantities
\be
\beta_a(h,g)=\alpha(a,h,g)\,\alpha(h, h^{-1} ah, g)^{-1}\,
 \alpha(h,g, (hg)^{-1} ahg ).
\label{beta}
\ee
It follows from this definition that the $\beta_a$'s are normalised twisted
cocycles
on $G$, namely they satisfy
\be
\beta_a(x,y) \, \beta_a(xy,z) = \beta_a(x,yz) \, \beta_{x^{-1}ax}(y,z),
\qquad \forall x,y,z \in G.
\label{cocy}
\ee
Furthermore, the restriction of each $\beta_a$ to $C_G(a)$ clearly is a
normalised
2-cocycle. As such they define projective representations of $C_G(a)$.

The primary fields $\Phi^\alpha$ in the model twisted by a
given 3-cocycle $\alpha$ will consist of all pairs $(a,\tilde \chi)$ where
$a \in \T$ (as before) and $\tilde \chi\in\beta_a$-Irr$(C_G(a))$.
As a consequence of the inequality $r(H,\beta)\le k(H)$, we find
\be
|\Phi^\alpha|=\sum_{a\in \T} \; r(C_G(a),\beta_a) \le |\Phi|.
\ee

A simplification seems to occur when the 3-cocycle $\alpha$ is such that
each 2-cocycle $\beta_a$ is a coboundary on $C_G(a)$, which is
the case  considered in \cite{DVVV}. 
We call the resulting twisting ``CT'' (cohomologically trivial). Note 
that when
$M(C_G(a))=0$ for all $a$, any twist $\alpha$ will automatically be CT. By no
means though
are CT twistings restricted to the groups whose centralisers have trivial
Schur multipliers --- see
\S6.2 for a class of examples.  A CT-twisted theory has the same number of
primaries
as the untwisted one considered in the previous sections (but different modular
matrices).

For CT twistings, \cite{DVVV,DW} 
proposed explicit formulae for the modular matrices that resemble
 the untwisted case except with additional phases. 
 Each $\beta_a$ being a coboundary on $C_G(a)$
by hypothesis, (\ref{cocy}) can be used to show\footnote{ We  thank 
Peter B\'antay for correspondence on this point.} that one can find 1-cochains
$\epsilon_a : C_G(a) \rightarrow U(1)$ for which $\epsilon_a(e)=1$ and
both
\bea
&&\beta_a(h,g) = (\delta\epsilon_a)(h,g) =
\epsilon_a(h)\,\epsilon_a(g)\,\epsilon_a(hg)^{-1},\label{epsilon1}\\
&&\epsilon_{x^{-1}ax}(x^{-1}hx)={\beta_a(x,x^{-1}hx)\over\beta_a(h,x)}
\epsilon_a(h),
\label{epsilon2}
\eea
for all $g,h\in C_G(a)$ and $x\in G$. Now, if $\tilde\rho$ is a 
$\beta_a$-representation with character $\tilde\chi$,
then
clearly $\rho(g) = \epsilon_a^{-1}(g) \tilde\rho(g)$ is a linear
representation with character $\chi = \epsilon_a^{-1} \tilde\chi$. The
modular matrices then become \cite{DVVV}
\bea
&& S^{\alpha}_{(a,\chi),(a',\chi')} = {1\over |C_G(a)|\cdot|C_G(a')|}
\sum_{g\in G(a,a')}\chi^*(ga'g^{-1})\,\chi'^*(g^{-1}ag)\,\sigma^*(a|ga'g^{-1}),
\hskip 1truecm \label{s}\\
&& T^{\alpha}_{(a,\chi),(a',\chi')}
= \delta_{a,a'}\delta_{\chi,\chi'}{\chi(a) \over  \chi(e)}\,\epsilon_a(a),
\label{t}
\eea
where the function $\sigma(\cdot|\cdot)$ is
\be
\sigma(h|g)=\epsilon_h(g)\,\epsilon_g(h).
\label{sigma}
\ee
(Since $g\in C_G(h)$ if and only if $h\in C_G(g)$, this definition makes
sense.) It is
easy to see that (\ref{s}),(\ref{t}) are (essentially, i.e. up to a
relabelling of the characters $\chi$) independent of the choice of 1-cochains
$\epsilon_a$ in (\ref{epsilon1}),(\ref{epsilon2}). These equations permit CT twists to be
analysed as
thoroughly as the untwisted data.

The interpretation involving 3-cocycles was developed in \cite{DW}.
Though they give
the fusion coefficients for arbitrary $\alpha$, neither they nor to
our knowledge anyone else has given $S^\alpha$ and $T^\alpha$ for
nonCT $\alpha$, {\it explicitly} in terms of quantities directly
associated with $G$ (the closest is
\cite{B3}, which gives $S^\alpha$ in terms of $D^\alpha(G)$
characters).  The formulae in the most general case can be
obtained  by going back to first principles.

The quantum double $D^{\alpha } (G) $ is a finite dimensional
 quasi-Hopf quasi triangular algebra. Its
representations form a  braided monoidal modular category with
 explicitly known
universal braiding morphisms $R_{12}$ ,  $R_{21}$. The point is that
 (up to normalisation)  the mapping
class group  representations, in our case the matrices $S$ and $T$, can
be identified with Markov traces of intertwiners defined from coloured
 ribbon links. In particular,
$S_{ij}$ corresponds to  the $i,j$ coloured Hopf link and $T_i $
 to the  twist $\theta$ (or $v^{-1} $ in the \cite{AC} convention).
 We then obtain (as explained in the Appendix)
\bea
S^{\alpha}_{(a,\tilde\chi),(b, \tilde\chi')} &=&
 S^{\alpha}_{(e,1),(e,1)} \  {\rm Tr}_{ (a,\tilde\chi),(b, \tilde\chi')}\
         \Big( R_{21}  R_{12} \Big)^*   \nonumber\\
 &=&  {1\over |G|}
     \sum_{g\in K_a  , g' \in K_b \cap C_G(g) }
     \Big(
{ \beta_{g}(g',x^{-1} ) \beta_{g'}(g,y^{-1} ) \over
  \beta_{g}(x^{-1} ,h )  \beta_{g'}(y^{-1} ,h') }   \Big)^* \
 \tilde\chi( h)^* \, \tilde\chi'(h' )^*, \nonumber \\
  &=&   {1\over |G|} \sum_{g\in K_a , g' \in K_b  \cap C_G(g) }
    \Big(
  { \beta_{a}( x ,g' )   \beta_{a}( xg' ,x^{-1}   ) \beta_{b}(y ,g )
 \beta_{b}( yg ,y^{-1}  )
  \over  \beta_{a}(x ,x^{-1} ) \beta_{b}(y , y^{-1} ) }\Big)^*  \
                 \tilde\chi( h)^* \, \tilde\chi'(h' )^*,\nonumber\\
\label{tws}
\eea
where $g = x^{-1} ax = y^{-1}h' y,\ g' = y^{-1} b y = x^{-1}h x,
\ h\in C_G(a),\  h'\in C_G(b)$, and
\be
T^{\alpha}_{(a,\tilde\chi),(b,\tilde\chi')}
 = \delta_{a,b}\delta_{\tilde\chi,\tilde\chi'}
{\tilde\chi(a)  \over \tilde\chi(e)}\,T^\alpha_{(e,1),(e,1)},
\label{twt}
\ee
where $T^\alpha_{(e,1),(e,1)}$ can equal any third root of 1 (i.e.\
$c$ is a multiple of 8), as for the untwisted data. The normalisation
$S^\alpha_{(e,1),(e,1)}={1\over |G|}$ can be obtained from the
orthogonality relations (\ref{orth}),(\ref{compl}) of projective characters.

The phases $\beta_g $ here are defined by (\ref{beta}). Note that they are
evaluated
in (\ref{tws}) on elements which are not in $C_G(a)$ or $C_G(b)$.
Unfortunately this makes the derivation of (\ref{s}) from (\ref{tws})
more difficult --- see question (1) in \S7.

If we let $n$ be the cohomological order of $\alpha$, then the
$\beta_a$ will also have $n$th roots of 1 as its values. Thm.\ 6.5.15
in \cite{K2} then implies that the projective $\beta_a$-characters
will have values in $\Q[\xi_{n\,e(G)}]$. Hence the entries of
$S^\alpha$ and $T^\alpha$ will also lie in that field. Note that this
field specialises to $\Q[\xi_{e(G)}]$ in the untwisted ($n=1$) case,
which was our previous result.

That $S^\alpha$ and $T^\alpha$ depend only on the cohomology class
$\alpha\in H^3(G,U(1))$, is clear from the $D^\alpha(G)$
interpretation. A direct derivation of this for CT $\alpha$ is
sketched in \cite{DW}.

\subsection{Analysis of twisted modular data}

It is important to realise that the 2-cocycle $\beta_e$ in (\ref{beta})
is identically 1. Thus $(e,\chi)\in\Phi^\alpha$ for any
$\chi\in{\rm Irr}(G)$, and we obtain the useful formula
\be
S^\alpha_{(e,\chi),(b,\tilde{\chi}')}={1\over
  |C_G(b)|}\tilde{\chi}'(e)\,{\chi}(b)^* \label{row0}
\ee
We immediately see from this that once again all quantum dimensions
will be integers, and $(a,\tilde{\chi})\in\Phi^\alpha(G)$ will be a
simple current iff $a\in Z(G)$ and $\tilde{\chi}$ is degree-one
(which implies $\beta_a$ is coboundary).

Rationality of the entries $S_{(e,1),(b,\tilde{\chi}')}$ implies all Galois
parities $\epsilon_\si(a,\tilde{\chi})=+1$, and also that the vacuum
$(e,1)$ is fixed by all Galois automorphisms, exactly as before.

{}From (\ref{row0}) we also learn about the Galois action on arbitrary
primaries. Choose $\ell\in\Z_{|G|^2}^\times$, then
$\si_\ell(a,\tilde{\chi})= (a_\ell,\tilde{\chi}_\ell)$, where
$a_\ell$ denotes the element in $K_{a^\ell}\cap \T$ as before, and where
$\tilde{\chi}_\ell$ is some projective $\beta_{a_\ell}$-character
with dimension equal to that of $\tilde{\chi}$. To see this, consider
\be
\chi(a_\ell)^*=
\si_\ell{S^\alpha_{(e,\chi),(a,\tilde{\chi})}\over
 S^\alpha_{(e,1),(a,\tilde{\chi})}} =\si_\ell\chi(a)^*=\chi(a^\ell)^*
\ee
for all $\chi\in{\rm Irr}(G)$, and hence $a_\ell\in K_{a^\ell}$.
In particular, specialising to $\ell=-1$ tells us about charge conjugation.

Proposition 1(a),(c),(e),(f) are thus exactly as before. As mentioned
before, the order of
$T^\alpha$ will divide $ne$, where $n$ is the order of $\alpha$, so in
particular the order of $T^\alpha$ will always divide $|G|^2$.
We also see directly from (\ref{tws}) that $S^{\alpha}_{(a,\tilde{\chi}),
(b,\tilde{\chi}')}=0$ unless $K_b\cap C_G(a)$ has $\beta_a$-regular elements
and $K_a\cap C_G(b)$ has $\beta_b$-regular elements. Thm.\ 1 becomes

\noindent{\bf Theorem 3.}
Let $S$ and $T$ be the Kac-Peterson matrices corresponding to an affine algebra
$X_r^{(1)}$ at some level $k \geq 1$ (where $X_r$ is simple). Let $G$ be a
finite
group with ${ S^\alpha}(G) = S$ and ${ T^\alpha}(G)=\varphi\,T$ for
some third root $\varphi$ of 1. Then either:

\begin{itemize}

\item[{\it (i)}] as before, either $(X_r,k)=(E_8,1)$ and $G=\{e\}$, or
 $(X_r,k)=(D_{8n},1)$ and $G=\Z_2$;

\item[{\it (ii)}] $(X_r,k)=(A_{n^2-1},1)$ and $G=\Z_n$, $n$ odd, for a
specific twist;

\item[{\it (iii)}] $(X_r,k)=(B_{(m^2-1)/2},2)$ and $G={\frak D}_m$, $m$
odd, for a
specific twist.
\end{itemize}

The proof is very similar to that of Theorem 1, the only difference
being the specific handling of the finitely many algebras and levels
which survive the Galois and quantum dimension arguments. For example,
use the formulae $S_{00}={1\over \sqrt{r+1}}$ and $c=r$ for
$A_r^{(1)}$ level  1. The demonstration of {\it (ii)} and {\it (iii)} is made
explicit in section 6.

We would also like  an analogue of Thm.2. This is more
difficult, but a key observation is
that $|\Phi^\alpha|\ge 2k(G)-1$. Indeed, $\beta_e\equiv 1$ gives us
$k(G)$ primaries of the form $(e,\chi)$, and the remaining $k(G)-1$
conjugacy classes $K_a$ will each contribute $r(a,\beta_a)\ge 1$
primaries. Also important is the observation made in \S6.2 that
$\beta_a$ will always be coboundary when $C_G(a)\cong \Z_n^2$ for some $n$.

\smallskip\noindent{\bf Theorem 4.} The only groups with at most 20
primaries are ${\c}_1$, ${\c}_2$, ${\c}_3$, ${\frak S}_3$,
${\frak A}_4$, $\Z_4$, $\Z_2\times\Z_2$, ${\frak D}_5$, ${\frak S}_4$,
${\frak D}_4$, and the order 48 Frobenius group $\Z_2^4\times_f\Z_3$
(defined in \cite{VL}), with at least
1, 4, 9, 8, 14, 16, 16, 16, 18, 19, and  19 primaries.
\smallskip

The proof is like that of Thm.2, using the tables of \cite{VL}.
For instance, we know it is sufficient to consider up to class number
$k(G)=8$. Consider e.g.\ $G={\frak S}_4$: it has $k(G)=5$ and
centralisers ${\frak S}_4$, ${\frak D}_4$, $\Z_4$,
$\Z_2\times\Z_2$, and $\Z_3$. Thus  a lower bound for $|\Phi^\alpha|$
is $5+2+4+4+3=18$.

In \S3.3 we gave a basis in terms of which the untwisted $S$ and $T$ become
permutation matrices. The analogue of this for the twisted data is as
follows \cite{B3}. Let $C^\alpha(G_{comm})$ denote the space of all
functions $f:G\times G\rightarrow \C$ for which $f(a,b)=0$ unless $a$
and $b$ commute, and which obey the formula
\be f(x^{-1}ax,x^{-1}bx)={\beta_a(x, x^{-1}bx)\over\beta_a(b,x)}f(a,b)\ee
Then $S$ and $T$ act on  $C^\alpha(G_{comm})$ by
$(Sf)(a,b)=\beta_b(a,a^{-1})^*f(b,a^{-1})$ and $(Tf)(a,b)=\beta_a(a,b)\,
f(a,ab)$. See Questions 5 and 7 of \S7. 
Incidentally, this action is completely natural in the case of twisted
    partition functions which involve projective representations of the
    symmetry group in some twisted sectors, and follows from the action
    of the modular group on the homological cycles of the torus --- see 
\cite{lrv}     for examples in WZW models.


\section{Twisted examples}

We give in this final section a few examples of twisted finite group
modular data,
in order to give a flavour as to how they differ from the untwisted ones.

\subsection{Abelian cyclic groups}

One can easily illustrate the previous formalism in the case of a cyclic group
$G=\Z_n $. Write  $\rest{\cdot}:\Z\rightarrow\{0,1,\ldots,n-1\}$ for
reduction modulo $n$.
One has $H^3(\Z_n,U(1)) = \Z_n$, and the following explicit representatives
\be
\alpha_q(g_1,g_2,g_3) = \exp{\Big\{2\i\pi q g_1(g_2 + g_3 - \rest{g_2 +
g_3})/n^2\Big\}},
\ee
where $q \in \Z_n$ parametrizes the different classes.

One easily computes that $\beta_a(h,g) = \alpha_q(a,h,g)$ is a coboundary for
every $q$ (expected since $M(\Z_n)=0$). One finds $\sigma(h|g) = e^{4\i\pi
qhg/n^2}$.

The linear characters of $\Z_n$ are $\chi_{\ell}(a) = e^{2\i\pi a\ell/n}$ for
$\ell \in \Z_n$, and the modular matrices take the simple forms
\bea
&& S^{\alpha_q}_{(a,\chi_\ell),(a',\chi_{\ell'})} = {1 \over n} \;
\exp{\Big\{-2\i\pi
[2qaa' + n(a\ell' + a'\ell)]/n^2\Big\}}, \label{szn} \\
&& T^{\alpha_q}_{(a,\chi_\ell),(a,\chi_{\ell})} = \exp{\Big\{2\i\pi
[qa^2 + na\ell]/n^2\Big\}}.
\eea
Note that the charge conjugation $C=S^2$ is given $C(0,\chi_\ell) =
(0,\chi_{-\ell})$ and $C(a,\chi_\ell) = (n-a,\chi_{-\ell-2q})$ if $a \neq 0$.

The identity corresponds to $(0,\chi_0)$, and as in the untwisted case, all
primary
fields are simple currents. The Verlinde formula yields the fusion coefficients
\be
N_{(a,\chi_\ell),(a',\chi_{\ell'})}^{(a'',\chi_{\ell''})} =
\delta_{a'',\rest{a+a'}} \, \delta_{\ell'',\ell + \ell' + 2q(a+a'-a'')/n}.
\label{fus}
\ee
This is in agreement with the charge conjugation, if one thinks of the
conjugate
of a field $\phi$ as the unique field $\phi^*$ such that the fusion $\phi
\times
\phi^*$ contains the identity. The fusion group (the group of simple
currents) is
isomorphic to $\Z_f \times \Z_{n^2/f}$ with $f = {\rm GCD}(2q,n)$.

The entries of $S$ and $T$ lie in $\Q(\xi_{n^2})$, which has Galois group
$\Z_{n^2}^\times$. It is not difficult to see that the Galois action on the
primaries is by fusion powers (as is always the case for simple currents):
\be
\sigma_h(a,\chi_\ell) = (a,\ell)^{\times h}, \qquad h \in \Z_{n^2}^\times.
\ee

The physical invariants are known in all cases, since all primaries are simple
currents. Their number varies much with the value of $q$ (of $f$). Two extreme
cases are $f=1$ (`maximal' twisting) for which the number of physical
invariants
is equal to $\sigma_0(n^2)$, the number of divisors of $n^2$ \cite{Deg},
and $f=n$
(no twisting), for which their number is $2(n+1)$ if $n$ is odd prime
\cite{KS}.

We close this simple example by showing that it gives, for a specific twisting,
the affine modular data of su$(n^2)$, level 1, if $n$ is an odd integer.

One may first make a few simple observations. The central charge of su$(n^2)_1$
is equal to $c \equiv n^2-1 \equiv 0$ (mod 8) for $n$ odd. The affine
quantum dimensions
$S_{0,j}/S_{0,0} = 1$ all equal 1, so all affine primaries are simple currents,
and form a group isomorphic to $\Z_{n^2}$.
Thus one may hope for a relation with
twisted $G=\Z_n$ data  where the twist obeys $f={\rm GCD}(q,n) = 1$.

The affine primaries can be labelled by integers $j=0,1,2,\ldots,n^2-1$ modulo
$n^2$. The affine matrices $S$ and $T$ are
\bea
&& S^{\rm aff}_{j,j'} = {1 \over n} \, e^{2\i\pi jj'/n^2}, \\
&& T^{\rm aff}_{j,j} = T_{0,0} \, e^{2\i\pi j(n^2-j)/2n^2},
\eea
where $T_{0,0} = e^{2\i\pi c/24}$ is some third root of unity, which
we will ignore.

It is now a simple matter to see that the two sets of matrices exactly coincide
provided one chooses the twisting parameter as $q={n^2-1 \over 2}$. The
bijection $j \leftrightarrow (a,\chi_\ell)$ between the two sets of primary
fields is given by $j = a - n \ell$. Note that for this specific value of
$q$, the
fusion coefficients (\ref{fus}) amount to the addition modulo $n^2$.

\subsection{Abelian non-cyclic groups}

The simplest non-cyclic group is $G=\Z_n^2$, but it leads to nothing really
new. The cohomology group $H^3(\Z_n^2,U(1)) = \Z_n^3$ has three generators,
but all
3-cocycles $\alpha$ lead to $\beta$'s which are all coboundaries. Thus all
twistings are CT, despite the fact that the Schur multiplier $M(\Z_n^2) = \Z_n$
is not trivial. (This fact was important for the proof of Thm.4.)

More interesting is the case $G = \Z_n^3$, for which $M(\Z_n^3)=\Z_n^3$ and
$H^3(\Z_n^3,U(1)) = \Z_n^7$. Following \cite{dWP,dwp}, the generators of $H^3$ can
be taken to be (same notations as above, the group elements are triplets
$a=(a_1,a_2,a_3)$)
\bea
&& \alpha_I^{(j)}(a,b,c) = \exp{\Big\{2\i\pi a_j(b_j + c_j - \rest{b_j +
c_j})/n^2\Big\}},\qquad 1 \leq j \leq 3,\\
&& \alpha_{II}^{(jk)}(a,b,c) = \exp{\Big\{2\i\pi a_j(b_k + c_k - \rest{b_k +
c_k})/n^2\Big\}},\qquad 1 \leq j < k \leq 3,\\
&& \alpha_{III}(a,b,c) = \exp{\Big\{2\i\pi a_1 b_2 c_3/n\Big\}}.
\eea

An arbitrary 3-cocycle is a monomial in the generators, but only those which
involve a non-trivial power of $\alpha_{III}$ define non-CT twistings.
In other words, all $\alpha$ which contain a fixed cocycle of type III
give rise to 2-cocycles $\beta_a$ which are cohomologically equivalent, and
hence lead to theories with the same number of primaries. In order to give a
first feeling for non-CT twistings, we will compute the number of primary
fields.
It is sufficient to take $\alpha$ of type III, namely
$\alpha = \alpha_{III}^q$, for $q \in \Z_n$. The 2-cocycles one obtains are
then
\be
\beta_a(b,c) = \exp{\Big\{2\i\pi q (a_1 b_2 c_3 - b_1 a_2 c_3 + b_1 c_2
a_3)/n\Big\}}.
\label{coc}
\ee

Given $a$, we want to count the number of classes $b$ (elements here) which are
$\beta_a$-regular, i.e. which satisfy $\beta_a(b,c) = \beta_a(c,b)$ for all
$c$. Taking successively $c=(1,0,0),(0,1,0)$ and $(0,0,1)$, the
$\beta_a$-regular elements $b$ are those which satisfy
\be
a_2 b_3 - a_3 b_2 \equiv a_1 b_3 - a_3 b_1 \equiv a_1 b_2 - a_2 b_1 \equiv
0\ ({\rm mod}\ f),
\label{reg}
\ee
where $f = n/{\rm GCD}(q,n)$. The number of solutions $(b_1,b_2,b_3) \in
\Z_n^3$ to this modular linear system is equal to $n^3 \cdot [{\rm
GCD}(a_1,a_2,a_3,f)/f]^2$, which is the result announced in (\ref{zn3}).

It remains to sum those numbers for all $a$ to obtain the number of primaries.
The result is an arithmetical function, best expressed in terms of the prime
decomposition of $f = \prod_p \, p^{k_p}$:
\be
|\Phi^\alpha| = {n^6 \over f^3} \prod_{p | f \atop p {\rm \ prime}}
\Big[(p^{k_p} - 1)(1 + p^{-1} + p^{-2}) + 1\Big].
\ee

The modular matrices can be given quite explicitly in the general case, for
all $n$
and for any type of 3-cocycle. However, they are complicated arithmetic
functions of
the various parameters, something that obscures the structure. To simplify, we
consider here the case when $n$ is an odd prime number, and when the
3-cocycle is
$\alpha_{III}$.

When $G$ is abelian, all factors in the formula (\ref{tws}) for $S^\alpha$ that
involve the cocycles drop out, and we are left with the simple expressions:
\be
S^{\alpha}_{(a,\tilde\chi),(b,\tilde\chi')} = {1\over |G|}
\tilde\chi^*(b) \, \tilde\chi'^*(a), \qquad
T^{\alpha}_{(a,\tilde\chi),(b,\tilde\chi')}
= \delta_{a,b} \, \delta_{\tilde\chi,\tilde\chi'} \, {\tilde\chi(a) \over
\tilde\chi(e)},
\ee
where $\tilde \chi$ and $\tilde \chi'$ are respectively $\beta_a$- and
$\beta_{b}$-projective characters, for the cocycles given above in
(\ref{coc}) with
$q=1$.

It remains to compute the projective characters. To simplify, consider
$n$ an odd prime. One then finds $n$
inequivalent irreducible $\beta_a$-projective representations of dimension
$n$ if $a$
is not the identity, while there are of course $n^3$ representations of
dimension 1 if
$a=e$. Depending on the value of $a=(a_1,a_2,a_3)$, the characters are
given in the
following table, where it is implicit that the element $g=(g_1,g_2,g_3)$
must be
$\beta_a$-regular for the character not to vanish. In the  first three
cases, the
character label $u$ runs over $\Z_n$, and in the last column, $\vec u$
takes all
values in $\Z_n^3$.

\medskip
\begin{center}
\begin{tabular}[t]{c|c|c|c|c}
 & $a_1 \neq 0$ & $a_1=0,\,a_2 \neq 0$ & $a_1=a_2=0,\,a_3 \neq 0$ &
$a_1=a_2=a_3=0$ \\
\hline
$\tilde \chi(g)$ & $n \, \xi_n^{a_1^{-1}ug_1 - a_1^{-1}a_2a_3g_1^2/2}$ & $n \,
\xi_n^{a_2^{-1}ug_2}$ & $n \, \xi_n^{a_3^{-1}ug_3}$ & $\xi_n^{\vec u \cdot
\vec g}$ \\
\end{tabular}
\end{center}
\medskip

The condition that $g$ must be $\beta_a$-regular makes the components of
$a$ and $g$ play a
symmetrical role. If $a_2$ is also invertible for instance, then
(\ref{reg}) yields
$a_1^{-1}g_1 = a_2^{-1}g_2$, so that the first character value is also
equal to $\tilde{\chi}(g) = n
\,\xi_n^{a_2^{-1}ug_2 - a_1a_2^{-1}a_3g_2^2/2}$.

The primary fields are thus $(e,\chi_{\vec u})$ and $(a,\tilde{\chi}_u)$, for a
total of
$|\Phi^\alpha| = n^3 + (n^3-1)n = n^4 + n^3 - n$.

The formulae for $S$ and $T$ are now straightforward to establish. Taking
the condition of
$\beta$-regularity into account, one finds that $S$ is almost block-diagonal:
\bea
&& \hspace{-10mm} S^\alpha_{(a,\tilde\chi_u),(b,\tilde\chi_{u'})} = \nonumber\\
&& \hspace{-10mm} {1 \over n} \pmatrix{
{1 \over n^2} & {1 \over n} \xi_n^{- u_1b_1 - u_2b_2 - u_3b_3} &
{1 \over n} \xi_n^{- u_2b_2 - u_3b_3} & {1 \over n} \xi_n^{- u_3b_3} \cr
{1 \over n} \xi_n^{- u'_1a_1 - u'_2a_2 - u'_3a_3} &
{\xi_n^{- ua_1^{-1}b_1 - u'b_1^{-1}a_1 + (a_2a_3b_1 + b_2b_3a_1)/2}
\atop \times \delta(b_2-a_1^{-1}a_2b_1) \, \delta(b_3-a_1^{-1}a_3b_1)} & 0
& 0 \cr
{1 \over n} \xi_n^{- u'_2a_2 - u'_3a_3} & 0 & {\xi_n^{- ua_2^{-1}b_2 -
u'b_2^{-1}a_2}
\atop \times \delta(b_3-a_2^{-1}a_3b_2)} & 0 \cr
{1 \over n} \xi_n^{- u'_3a_3} & 0 & 0 & \xi_n^{- ua_3^{-1}b_3 -
u'b_3^{-1}a_3}},\nonumber\\
\eea
where the blocks correspond to the subsets $\{a=e\}$, $\{a_1 \neq 0\}$ ,
$\{a_1=0,\;a_2
\neq 0\}$, and $\{a_1=a_2=0$, $a_3 \neq 0\}$.

The $T$ matrix is particularly simple
\be
T_{(a,\tilde\chi),(a,\tilde\chi)} = \cases{
1 & for $a=e=(0,0,0)$, \cr
\xi_n^{u-a_1a_2a_3/2} & otherwise.}
\ee

One may check that they satisfy the expected relations $S^2 = (ST)^3 = C$
with the charge
conjugation given by $C(e,\chi_{\vec u}) = (e,\chi_{-\vec u})$ and
$C(a,\tilde\chi_u) =
(a^{-1},\tilde\chi_{u-a_1a_2a_3})$ for $a \neq e$. More generally, the
Galois transformations
on the primary fields take a somewhat unusual form\footnote{The unusual
factor $\ell^2$ in
front of $u$ (instead of $\ell$) is a consequence of the insertion of
factors $a_i^{-1}$ in
front of $u$ in the projective characters. Those insertions are purely
conventional and help
make the symmetry in the components $a_i$ manifest.}
\be
\sigma_\ell(a,\tilde\chi_u) = \cases{
(e,\chi_{\ell\vec u}) & for $a=e$,\cr
\noalign{\smallskip}
(a^\ell,\tilde\chi_{\ell^2 u + a_1a_2a_3\ell^2(\ell-1)/2}) & for $a \neq e$,}
\qquad \ell \in \Z_n^\times.
\ee
All Galois parities are equal to $+1$. One also checks the relation
$\sigma_{\ell^2}T_{(a,\tilde\chi),(a,\tilde\chi)} =
T_{\sigma_\ell(a,\tilde\chi),\sigma_\ell(a,\tilde\chi)}$.

Finally the fusion algebra can be computed, which shows a structure
radically different
from the untwisted or CT case. Here too, the fusion coefficients
generically involve
the group law and the structure constants for the irreducible characters.
Since the
latter have degree 1 or $n$, the fusion of two primary fields
may contain 1, $n$ or $n^2$ primary fields, with possible multiplicities
(they turn
out to be 1 or $n$ only). The resulting formulae are tedious to write in full
form as various cases need be distinguished. As an illustration, we give the
coefficients when $a,b,c \neq e$:
\be
N_{(a,\tilde\chi_u),(b,\tilde\chi_{u'})}^{(c,\tilde\chi_{u''})} =
\delta(a+b-c) \cases{
1 \quad \hbox{for $c_1=0$ and $a_1,b_1 \neq 0$,} & \cr
\Big[n\,\delta(a \hbox{ is $\beta_b$-reg})\,\delta(a_1^{-1}u +
b_1^{-1}u' + a_2b_3 - c_1^{-1}u'')  & \cr
\hspace{1cm} + \delta(a \hbox{ is not $\beta_b$-reg})\Big] \quad \hbox{for
$c_1 \neq 0$ and
$a_1,b_1$ not both 0,} & \cr
\Big[n\,\delta(a \hbox{ is $\beta_b$-reg})\,\delta(a_2^{-1}u +
b_2^{-1}u' - c_2^{-1}u'')  & \cr
\hspace{1cm} + \delta(a \hbox{ is not $\beta_b$-reg})\Big] \quad \hbox{for
$a_1,b_1=0$ and
$a_2,b_2 \neq 0$,} & \cr
\Big[n\,\delta(a \hbox{ is $\beta_b$-reg})\,\delta(a_3^{-1}u +
b_3^{-1}u' - c_3^{-1}u'')  & \cr
\hspace{1cm} + \delta(a \hbox{ is not $\beta_b$-reg})\Big] \quad \hbox{for
$a_1,a_2,b_1,b_2=0$ and $a_3,b_3 \neq 0$.} & }
\ee

An intriguing observation made in \cite{dwp} (what he called {\it
  electric/magnetic duality}) is that the modular data for $\Z_2^3$
  twisted by $\alpha_{III}$ {\it equals} that of untwisted ${\frak
  D}_4$ (for an appropriate identification of primary fields), while
  the  twist $\alpha_I^{(1)}\alpha_{III}$ yields the ${\frak Q}_4$
  modular data. We will address this again in Question {\bf (8)} in \S7.

\subsection{Odd dihedral groups}

The simplest non-abelian groups are the dihedral groups $\Dm$. Their Schur
multipliers are equal to $0$ or $\Z_2$ for $m$ odd or even respectively.
For $m$
odd, all centralisers of elements of $\Dm$ also have trivial Schur multipliers,
implying that all twistings of the modular data will be CT. Despite that
fact, we
will precisely consider $m$ odd here\footnote{The specific case $m=3$ has been
treated in \cite{DVVV}.}, since we want to show that a particular CT
twisting of
the $\Dm$ modular data yields nothing but the affine modular data of the odd
orthogonal series $B_{\ell}$ at level 2, and for $m = \sqrt{2\ell+1}$.

The relevant group theoretic data for $\Dm$ have been recalled in section
3.2. The
number of primary fields is the same for all twistings, and given by $|\Phi| =
{m^2+7 \over 2}$.

The third cohomology group $H^3(\Dm,U(1)) = \Z_{2m}$ is cyclic and so all
3-cocycles are powers of some generator. They have been very explicitly
determined
in \cite{dWP}. Write the elements of $\Dm$ as $g = s^A\,r^a$ for $A=0,1$ and
$a=0,1,\ldots,m-1$. The group law then takes the form
\be
(A,a)(B,b) = (\rest{A+B}_2, \rest{(-1)^Ba + b}_m)
\ee
where the notation $\rest{x}_n$ means taking the residue of $x$ modulo $n$
between
0 and $n-1$. Then the 3-cocycles are given by \cite{dWP}
\be
\alpha((A,a),(B,b),(C,c)) = {\rm exp}\left\{-{2\i \pi p \over m^2} \Big[
(-1)^{B+C}a[(-1)^Cb + c - \rest{(-1)^Cb + c}_m] + {m^2 \over 2} ABC
\Big]\right\},
\ee
where $p=0,1,2,\ldots,2m-1$ labels the cohomology classes.

We now proceed to compute the matrices $S$ and $T$ from the formulas
(\ref{s}) and
(\ref{t}). We first compute
\be
G(e,g) = \Dm, \quad G(r^k,r^l) = \Dm, \quad G(s,s) = \{e,s\}, \quad G(s,r^a) =
\emptyset.
\ee
In particular, the last equality implies
$S^\alpha_{(s,\chi),(r^k,\chi')}=0$. The
remaining entries of $S^\alpha$ only require knowing the following values of
$\sigma$, which easily follow from (\ref{beta}), (\ref{epsilon1}) and
(\ref{sigma}),
\be
\sigma(e|g) = 1, \qquad \sigma(s|s) = (-1)^p, \qquad
\sigma(r^a|r^b) = {\rm exp}\left({-4\i \pi pab \over m^2}\right).
\ee

The rest is just a matter of a few calculations. We will write the two modular
matrices with their rows and columns indexed by
\be
(e,\psi_0), (e,\psi_1), (s,1), (s,\chi\neq 1), (e,\chi_i), (r^k,\chi_\gamma),
\ee
where $\chi_\gamma$ are characters of the group $\Z_m$, with values
$\chi_\gamma(a) = {\rm e}^{-2\i\pi a\gamma/m}$. The indices $i$ and
$\gamma$ run
from 1 to $m-1 \over 2$ and $m$ respectively.

One finds
\be
S^\alpha = {1 \over 2m} \; \pmatrix{1 & 1 & m & m & 2 & 2 \cr
1 & 1 & -m & -m & 2 & 2 \cr
m & -m & (-1)^p m & (-1)^{p+1} m & 0 & 0 \cr
m & -m & (-1)^{p+1} m & (-1)^p m & 0 & 0 \cr
2 & 2 & 0 & 0 & 4 & 4\cos{2\pi i\gamma \over m} \cr
2 & 2 & 0 & 0 & 4\cos{2\pi i\gamma \over m} & 4\cos{2\pi (2pkk'+\gamma k' +
\gamma' k) \over m^2}},
\ee
and
\be
T^\alpha = {\rm diag}\Big[1,1,{\rm e}^{4\i\pi p/8},-{\rm e}^{4\i\pi p/8},1,{\rm
e}^{2\i\pi(pk^2-mk\gamma)/ m^2}\Big].
\ee

\medskip
On the affine side, the alc\^ove of $B_{\ell}$ level 2 contains $\ell + 4$ primary
fields, corresponding to the weights: 0, $2\omega^1$, $\omega^1 +
\omega^\ell$,
$\omega^\ell$ and $\nu^j$, defined as $\nu^j = \omega^j$ for $1 \leq j \leq
\ell-1$, and $\nu^\ell = 2\omega^\ell$. The conformal central charge is
equal to
$2\ell$, which is multiple of 8 if $\ell$ is a multiple of 4. For
simplicity, we
will use the height variable, $n=2\ell+1$.

The affine $S$ matrix, with the primary fields labelled as above, is equal to
\cite{kw}
\be
S = {1 \over 2\sqrt{n}} \; \pmatrix{1 & 1 & \sqrt{n} & \sqrt{n} & 2 \cr
1 & 1 & -\sqrt{n} & -\sqrt{n} & 2 \cr
\sqrt{n} & -\sqrt{n} & \sqrt{n} & -\sqrt{n} & 0 \cr
\sqrt{n} & -\sqrt{n} & -\sqrt{n} & \sqrt{n} & 0 \cr
2 & 2 & 0 & 0 & 4\cos{2\pi jk \over n}},
\ee
while the affine $T$ is
\be
T = T_{0,0} \cdot {\rm diag}\Big[1, 1, -{\rm e}^{2\i\pi\ell/8}, {\rm
e}^{2\i\pi\ell/8}, {\rm e}^{\i\pi j(n-j)/n}\Big].
\ee

The comparison of the two pictures is now easy. As far as the first four
primary
fields are concerned, the two sets of modular data coincide provided
$m=\sqrt{n}=
\sqrt{2\ell+1}$ and $p \equiv 0$ (mod 2) (the individual identifications
among the
third and fourth fields depend on the value of $\ell$ (mod 8)). It implies
that the
finite group and the affine theories have the same number of primaries,
$\ell+4 =
{m^2+7
\over 2}$.

The remaining $\ell$ fields, on the affine side, are the $\nu^j$. They can
be set
in correspondence with the finite group primary fields $(e,\chi_i)$ and
$(r^k,\chi_\gamma)$, depending on whether $j$ is a multiple of $m$ or not,
via the
following formula:
\be
\nu^{im} \sim (e,\chi_i), \qquad \nu^{k+m\gamma} \sim
(r^k,\chi_\gamma).
\ee
When $i,k$ and $\gamma$ run over their domain, the index $j$ indeed takes all
integer values from 1 to $\ell={n-1 \over 2}$.

One may then check that the two sets of modular matrices coincide for a
specific
choice of $p$. For $j=i\sqrt{n}$, one has $j(n-j)\equiv 0$ (mod $2n$),
while for
$j=k+\sqrt{n}\gamma$, one finds
\be
j(n-j) \equiv {\textstyle {n-1 \over 2}} k^2-k\gamma\sqrt{n} \equiv
pk^2 - mk\gamma\ ({\rm mod}\ 2n),
\ee
provided one makes the choice $p\equiv {n-1 \over 2}$ (mod $2n$). It gives
the unique
twisting for which the finite group modular data and the affine data
coincide. That
the $S$ matrices are equal is done in the same way.

Hence infinitely many physical invariant classifications for CT-twisted
nonabelian $G$ were done in \cite{G}. If we let $d = \sigma_0(m^2)-1$
denote the
number of divisors $d'\le m$ of $m^2$, then the number of physical invariants
of ${\frak D}_m$ for the given twist $\alpha$  is $d(d+3)/2+4$. Since
${\frak D}_3\cong{\frak S}_3$, we get `only' 9 physical invariants for
twisted ${\frak S}_3$ -- still a large number considering the small
number of primaries, but dwarfed by the 32 ones for the untwisted data.

\section{Questions}

We conclude by collecting a small number of the questions raised in
this paper. 
\smallskip

\noindent{\bf (1)}\footnote{ We thank Peter B\'antay  for  correspondence regarding this point.} 
We have been unable to derive the CT formula
(\ref{s}), appearing in e.g.\ \cite{DVVV}, from our general formula (\ref{tws}),
except in special cases such as when $G$ is abelian. Though this isn't
strictly necessary, it would make a nice
consistency check of the modular data. 
Our formula (\ref{tws}) for $S^\alpha$ has the strength
that it is given explicitly in terms of quantities directly associated
with $G$, but it has the weakness that it involves the 2-cocycles $\beta_a$
away from $C_G(a)$, and this is what complicates our attempt to derive (\ref{s}).
The derivations in \cite{DW,B3} involve considering the $\beta_a$'s
at the centralisers only.

\smallskip\noindent{\bf (2)} It should be possible to generalise the
observations in \S4.2 by considering the relations of the modular data
for $G$ with that of normal subgroups and quotient groups. This should
yield an analogue of ``conformal embeddings'' for this data.

\smallskip\noindent{\bf (3)} Can we recover the character table from the
matrices $S$ and $T$?
Do non-isomorphic groups have different $S$ and $T$? These are big questions.
Probably the answer to the first is yes, and to the second is no. It turns out
to be very difficult to identify a group from its character table together with
nice additional information -- perhaps the modular data provides a means?
A natural place to look for nonisomorphic groups with identical modular
data are the order 16 groups with identical group algebras, or Brauer
pairs (i.e.\ groups with identical character tables and identical
power maps $K_a\mapsto K_{a^n}$).\footnote{ We thank John McKay for
  bringing Brauer pairs to our attention.} Perhaps
relevant is \cite{da}, which proves that two groups have the same
character table iff their group algebras are isomorphic as quasi-Hopf
algebras.

\smallskip\noindent{\bf (4)} We see in places a tantalising hint of
some sort of duality between the group element component of the
primary field, and its character component (see e.g.\ \S4.2). Is there any
way to make
this precise?

\smallskip\noindent{\bf (5)} Does the  SL$_2(\Z)$ representation
obtained from the finite group modular data,   factor through a
congruence subgroup? Presumably the answer is always yes. Certainly it
is true for untwisted data. The easiest way to see this uses the
permutation-like representation of \cite{DVVV,B3,KSSB}, so a natural starting point for the
twisted data would involve the basis given at the end of \S5.3. 

\smallskip\noindent{\bf (6)} An important fact for affine modular data
is the Kac-Peterson formula, which provides a useful interpretation for the
eigenvalues
$S_{\lambda\mu}/S_{0\mu}$ of the fusion matrices. It would be highly
desirable to find out what form if any that takes here.

\smallskip\noindent{\bf (7)} What is the analogue for the affine
modular data of the basis discussed at the end of \S5.3, in which $S$
and $T$ appear as generalised permutation matrices? 

\smallskip\noindent{\bf (8)} As mentioned at the end of \S6.2, twisted 
$\Z_2^3$ modular data yields \cite{dwp} that of untwisted ${\frak D}_4$ and 
${\frak Q}_4$ modular data. It is natural to expect that there are many more
such examples: e.g.\ circumstantial evidence (such as quantum
dimensions and numbers of primaries) suggests that appropriately twisted $\Z_p^3$ would yield the
modular data for the Heisenberg group (consisting of all
$3\times 3$ upper triangular matrices with 1's down the diagonal and
entries in $\Z_p$), or the order $p^3$ generalisation of $\frak{D}_4$ 
(presentation 
$\langle
a,b\,|\,a^{p^2}=b^p=b^{-1}aba^{-1-p}=e\rangle$). 
This is probably the analogue of ``rank-level duality'' 
for this finite group data. Note that it is not a coincidence that
$\Z_2^3$ and ${\frak D}_4$ both have the same order --- if the
matrices $S$ are to be equal, their entries $S_{0,0}$ must agree and
hence the two groups must have equal order. Again, perhaps relevant is \cite{da},
which may supply a starting point for a general theory.\smallskip

\bigskip\noindent{\bf Acknowledgements.} T.G.\ thanks F.\ A.\ Bais, 
P.\ B\'antay, J.\ McKay, D.\ Stanley, M.\
Walton, and B.\ Westbury; his research was supported in part by NSERC.
A.\ C.\ thanks D.\ Altschuler, J.\ Lascoux, V.\ Pasquier, J.B.\ Zuber. This
paper was initiated  at Oberwolfach and partially written at IHES, and we
thank both for their hospitality.


\bigskip
\appendix
\section{Appendix}

We sketch here  the derivation of formula (\ref{twt}) for $T^\alpha$.
 A basis of $D^\alpha(G)$ is labelled by pairs $(g,x)$ of group elements
 with algebra product
\be  (g,x) \cdot (k,y)\ :=  \delta_{g, xkx^{-1} } \
     \beta_g ( x,y) \  (g, xy).
\ee
Now, $D^\alpha(G)$ is  semi-simple and finite-dimensional over $\C $, so its
irreducibles $R(a,\tilde{\rho})$ are subrepresentations of the regular one.
They can be described as follows \cite{DPR}.
The representation space for $R_{(a,\tilde{\rho})}$ is
       spanned by vectors $ |x_j\rangle \otimes |w\rangle $,
and the action is give by
\be
R_{( a, \tilde{\rho } ) } (g,x) |x_j\rangle \otimes |w\rangle
= \delta_{g, x_k   a x_k^{-1} }\
   { \beta_g (x,x_j )  \over  \beta_g (x_k , h) } \
    |x_j
\rangle \otimes  \   \tilde\rho (h)  |w\rangle ,
\ee
where
$x_j$ are coset representatives for $G/C_G (a)$  and $( x_k , h) \in
G/C_G (a) \times C_G
(a) $ are uniquely defined by $ x\ x_j = x_k h $.
Here $\tilde\rho$
denotes a projective irreducible $\beta_a $-representation
 of $C_G (a) $ obtained by restriction of the regular representation
  of $D^{\alpha } (G) $.

        Contact with the regular representation has been made by
 Altschuler and Coste \cite{AC}.
        Let us give here an account of it.
        The central element $v^{-1} = \sum_{k\in G} (k,k) $
     is easily diagonalised:
     For each conjuguacy class $K_a $ of $G$ denote by $e(a) $
     the common order of
      its elements, then
\be
\omega_a := \prod_{j=0}^{e(a)-1} \
             \alpha  (g, g^j x,x^{-1}gx )
\ee
        is independent of the choice of $g \ \in K_a $ and $x \in G. $

        Then all eigenvalues of $ v^{-1} $ are the $e(a)$-th roots of
        the $ \omega_a $'s. For each such root $\lambda $, each
        $  g\in K_a $ and each class $ \{ g^j x \} $ of $G/\langle
        g\rangle $ , an eigenvector $\psi_{  \lambda }[g,x] $
        is easily constructed.

 That these eigenvalues of projective $\beta_a $-representations
  are given in terms of (roots of) values of a 3-cocycle raises the
interesting
  group theoretic question: is this a clever way of classifying or
  putting together all projective representations of a finite group
  in terms of 3-cocycles for extensions inside which it is a
  centraliser ?

           The central element $v$ is a constant
           on each $R_{( a, \tilde\rho ) } $,
        for convenience let us rather focus on $v^{-1} $, which has a simpler
expression and on its eigenvalues: they are necessarily $e(a)$-th roots of the
 important quantities $\omega_a$,
 numbers which should be considered
 as tabulation data of any 3-cocycle. All of their $e(a)$-th roots are
eigenvalues
         of $v^{-1} $.

 The following nice identity satisfied by the twisted non abelian
   2-cocycle and  valid for any $a,x \in G$, is  due to Altschuler:
\be
  \beta_{xax^{-1} } ( xax^{-1},\  x) =
    \beta_{xax^{-1} } (x,\ a) = \alpha   ( xax^{-1}, x,a)
\ee
  and is proved by direct evaluation in terms of the 3-cocycle $\alpha$.
 From it we get the remarkably simple expression:
    $R_{(a,\tilde\rho)} ( v^{-1} )\  \  |x_j\rangle \otimes |w\rangle =
        \  |x_j\rangle \otimes  \  \tilde\rho (a) |w\rangle = \lambda   \
         | x_j\rangle \otimes |w\rangle    $
        (here $|w\rangle $ spans the representation space of
       $ \tilde\rho $.
 Because of the expression
of  $v^{-1} $ the quotient of cocycles which appear in
$R_{(a,\tilde\rho)} $
  becomes here independent of $x= x_j = x_k $ and in fact equal to $1$.

  Since for $h \in C_G (a) $,
\be
\beta_{a} ( h,\ a ) = \beta_{a} (a ,\ h )= \alpha ( a,h,a)
\ee
   we easily get, using an induction on $j$ for $h= a^j $:
   $\tilde\alpha (a)^{e(a) } = \omega_a I$,
    $   \tilde\alpha (a) $ central and
   equal to $\lambda   I$.


\end{document}